\documentclass[11pt]{article}
\pdfoutput=1
\usepackage{jheppub}
\usepackage{float, extarrows, tikz-cd}
\usepackage{graphicx}
\usepackage{slashed}
\usepackage{tabularx,ragged2e}
\usepackage{amssymb}
\usepackage{subcaption}
\usepackage{amsmath,amssymb}
\usepackage{slashed}
\usepackage{caption}
\usepackage{xcolor}
\usepackage{dsfont}
\usepackage{verbatim}
\usepackage{mathtools, xcolor,ytableau, amsfonts,tikz}
\usepackage{graphicx}
\usepackage{physics}
\usepackage{simpler-wick}
\newcolumntype{C}{>{\Centering\arraybackslash}X}

\numberwithin{equation}{section}

\newcommand{\mL}{\mathcal{L}}
\newcommand{\mO}{\mathcal{O}}

\newcommand{\pd}{\partial}

\def\<{\langle}
\def\>{\rangle}

\title{Spontaneous symmetry breaking on surface defects} 
\author[a,b]{Gabriel Cuomo,}   
\author[c]{Shuyu Zhang}     
     
\affiliation[a]{Simons Center for Geometry and Physics, SUNY, Stony Brook, NY 11794, USA}      
\affiliation[b]{C. N. Yang Institute for Theoretical Physics, Stony Brook University, Stony Brook, NY 11794, USA}     
\affiliation[c]{Department of Physics and Astronomy, Stony Brook University, Stony Brook, NY 11794, USA}                             
           
\emailAdd{gcuomo@scgp.stonybrook.edu}					 
     \emailAdd{shuyu.zhang.1@stonybrook.edu}
                                           
\abstract{Coleman's theorem states that continuous internal symmetries cannot be spontaneously broken in two-dimensional quantum field theories (QFTs).  In this work we consider surface (i.e. two-dimensional) defects in $d$-dimensional conformal field theories (CFTs) invariant under a continuous internal symmetry group $G$.  We study under which conditions it is possible for a surface defect to break spontaneously a continuous internal symmetry.  We find that spontaneous symmetry breaking (SSB) is impossible under reasonable assumptions on the defect Renormalization Group (RG) flow. Counterexamples are possible only for exotic RG flows,  that do not terminate at a fixed-point. We discuss an example of this kind.  We also illustrate our no-go result with an effective field theory analysis of generic defect RG flows.  We find a generic \emph{weakly coupled} defect universality class (with no SSB),  where correlation functions decay logarithmically. Our analysis generalizes the recent discovery by Metlitski of the extraordinary-log boundary universality class in the $O(N)$ model.  }

\begin{document} 
\maketitle

\section{Introduction and summary}

Defects play several roles in Quantum Field Theory (QFT).   First,  they describe (point-like or extended) impurities and boundaries in quantum and statistical systems. Second,  topological defect operators are interpreted as (potentially non-invertible) symmetry generators, see \cite{Gaiotto:2014kfa,Chang:2018iay} and references therein. Finally, sometimes defects provide useful order parameters; for instance, the expectation value of the Wilson loop is a diagnosis of confinement \cite{Wilson:1974sk}.

In relativistic QFT, we can think of a defect in two equivalent ways.  On the one hand, a defect can be regarded as a nonlocal operator acting on the vacuum of the theory.  Alternatively, we can think of an extended operator as a modification of the Hamiltonian (and thus of the action) of the system in some region of space.  The system defined by the defect and the bulk QFT is often referred to as a defect QFT (DQFT).

Even when the bulk is conformal, it is known that defects still undergo nontrivial Renormalization Group (RG) flows.   As for bulk RG flows \cite{Zamolodchikov:1986gt,Komargodski:2011vj,Casini:2012ei}, one expects the defect RG flow to be irreversible. Recently, this has been established for local unitary defects of dimension $p\leq 4$ \cite{Friedan:2003yc,Jensen:2015swa,Wang:2021mdq,Cuomo:2021rkm,Casini:2023kyj} (see also \cite{Affleck:1991tk,Yamaguchi:2002pa,Gaiotto:2014gha,Casini:2016fgb,Casini:2018nym,Kobayashi:2018lil,Wang:2020xkc,Casini:2022bsu}).\footnote{The proofs of irreversibilty for $p\geq 2$ assume that the endpoint of the defect RG flow is described by a DCFT; as we comment below, this is not always the case. There are however no counterexamples to the irreversibility of the defect RG.}

At large distance,  one expects to typically find a critical defect operator.  This means that the DQFT becomes scale invariant at large distances. In most cases, scale invariance is enhanced to full conformal invariance \cite{Nakayama:2012ed}.
When both the bulk theory and the defect are conformal, the DQFT defined by the insertion of a $p$-dimensional defect preserves a\footnote{Sometimes, the transverse rotations $SO(d-p)$ may not be preserved by a conformal defect \cite{Aharony:2022ntz,Gabai:2022mya,Gabai:2022vri,Aharony_to_appear}; this will not be important for our analysis.}
\begin{equation}
SO(p+1,1)\times SO(d-p)
\end{equation}
subgroup of the $d$-dimensional conformal symmetry group $SO(d+1,1)$.  This defines a Defect Conformal Field Theory (DCFT) \cite{Liendo:2012hy,Billo:2016cpy}.  

In CFT, most DQFTs become scale (and conformal) invariant at large distances. However, this is not always the case.  Exceptions to this behaviour are known in free theories, in which case there exist defects that exhibit runaway RG flows \cite{Cuomo:2021kfm}. These are characterized by the lack of scale invariance at arbitrary large distances from the defect; for instance, one-point functions of bulk operators exhibit slower decay with the distance from the defect than in DCFTs \cite{Cuomo:2022xgw}.  As we will explain, these exotic RG flows are expected to be possible only in theories with moduli spaces of vacua.

In this work, we will be interested in surface, i.e. two-dimensional, defects in conformal field theories (CFTs).  Important applications of this setup include boundaries and interfaces in three-dimensional critical systems \cite{Diehl:1996kd,Cardy:2004hm},  
as well as supersymmetric surface operators in superconformal gauge theories \cite{Gukov:2006jk,Lunin:2007ab,Drukker:2008wr}.  Recently, the bootstrap program achieved significant progress in the study of conformal surface defects in different theories (see e.g. \cite{Liendo:2012hy,Gliozzi:2015qsa,Drukker:2020atp,Lauria:2020emq,Behan:2021tcn,Padayasi:2021sik,Herzog:2022jqv}).

In ordinary two dimensional QFTs (without defects), an important result is Coleman's theorem \cite{Coleman:1973aa}, that states that continuous internal symmetries cannot be spontaneously broken.  Coleman's theorem is the continuous version of the Mermin-Wagner-Hohenberg theorem for the quantum $O(N)$ model on a two-dimensional lattice at finite temperature \cite{Hohenberg,Mermin:1966fe} (see \cite{2001JPCM...13R.505G} for a survey on generalizations to different lattice systems). It is natural to ask under which conditions an analogous result holds for surface defects.

In this work we study under which conditions it is possible for continuous internal symmetries to be broken on surface defects in $d>2$-dimensional CFTs.\footnote{Notice that when the bulk theory is gapped one can integrate out the bulk degrees of freedom. Thus,  the IR theory is a local two-dimensional QFT on the defect and Coleman's theorem applies straightforwardly to surface defects in bulk gapped theories.} By spontaneous symmetry breaking we mean that one or more (defect or bulk) operators transforming in a nontrivial representation of a continuous internal symmetry $G$ acquire an expectation value, even if the defect \emph{preserves} the symmetry. It is important to distinguish this setup from defects whose boundary conditions \emph{explicitly} break the internal symmetry.\footnote{As a concrete example of the distinction between a defect breaking and one preserving the internal symmetry,  let us consider the normal and the extra-ordinary boundary conditions for the $d-$dimensional $O(N)$ model on the half plane $z>0$. Intuitively, the normal boundary breaks explicitly the symmetry to $O(N-1)$ by demanding that the fundamental field $\phi_A$ close to the boundary behaves as
\begin{equation}\label{eq_foot_normal}
\phi_A\sim a_{\phi}\frac{\delta_A^1}{z^{\Delta}}\,,
\end{equation}
where $z$ denotes the distance from the boundary, $a_{\phi}$ is an $O(1)$ number (the bulk-to-defect OPE coefficient of the identity) and $\Delta_{\phi}$ is the scaling dimension of $\phi_A$.  The condition \eqref{eq_foot_normal} for instance arises integrating $\phi_1$ on the boundary.  The extra-ordinary boundary class instead is less restrictive and demands
\begin{equation}\label{eq_foot_extra}
\phi_A\sim a_{\phi}\frac{n_A(y)}{z^{\Delta}}\,,
\end{equation}  
where $n_A(y)$ is an arbitrary field configuration on the boundary such that $n_A n_A=1$; as we will see, we can think of the components of the field $n_A(y)$ as the defect Goldstones.  The boundary class~\eqref{eq_foot_extra} can be concretely realized introducing a negative $O(N)$-invariant mass term $-m^2\int d^{d-1}y\,\phi_A\phi_A$ at $z=0$. For $N=1$ the two boundary conditions were shown to be effectively equivalent \cite{1994PhRvB..50.3894B}.}
We also demand that the expectation value of all bulk operators vanishes at infinite distance from the defect.\footnote{In theories with moduli spaces it is also possible to induce symmetry breaking in the bulk introducing a potential on the defect, see e.g. \cite{Prochazka:2020vog} for examples in a free scalar theory.  In this work we focus on the setup in which the vanishing boundary conditions at infinity preserve the symmetry.}

Coleman's theorem in QFT is a consequence of Goldstone's theorem, that implies the existence of massless particles because of spontaneous symmetry breaking. As well known, the propagator of a massless field in two-dimensions is a logarithm in position space; this implies that connected correlation functions of the order parameter grow (in absolute value) with the distance between the operators. Physically, this means that small local perturbations can affect the system at arbitrary large distance from the perturbation, and thus the ordered phase is unstable to quantum fluctuations.

A priori, it is not obvious if a constraint similar to Coleman's theorem should exist for defects in CFTs. Indeed,  the gapless bulk environment mediates non-local interactions between the Goldstone modes on the defect.  It was already pointed out in the original work by Mermin and Wagner that symmetry breaking might occur in generalizations of the two-dimensional Heisenberg lattice model with nonlocal interactions \cite{Mermin:1966fe}. This is because a nonlocal intereaction, in general, modifies the correlation functions of the Goldstones and may cure the pathological logarithmic behaviour at large distances. 

Despite the potential skepticisms raised by these observations, in \cite{Metlitski:2020cqy} it was shown that symmetric boundaries in the $3d$ critical $O(N)$ model never lead to symmetry breaking. Interestingly, for sufficiently small $N$ there still exist a \emph{weakly-coupled} Goldstone sector on the defect, despite the absence of SSB, and correlation functions decay logarithmically. This behaviour is special of two-dimensional boundaries. For $d>3$ the $O(N)$ model admits the \emph{extra-ordinary} boundary universality class \cite{PhysRevB.12.3885}, for which there exists a defect order parameter spontaneously breaking the symmetry.  The result of \cite{Metlitski:2020cqy} is thus nontrivial and motivates a more general investigation of the issue of symmetry breaking on surface defects.

The main results of this paper are the following
\begin{itemize}
\item We argue that no symmetry breaking can occur in unitary theories if the DQFT flows to a scale invariant fixed-point at energies below the would-be symmetry breaking scale.  We justify this claim both with a non-perturbative argument and with an explicit \emph{effective field theory} (EFT) analysis. The non-perturbative argument relies on the combination of Ward identities with the constraints imposed by the infrared scale invariance.\footnote{More precisely, we assume that the low energy theory admits a diagonalizable dilation operator with discrete spectrum.}  The EFT analysis consists in the study of the perturbative RG flow in two-dimensional non-linear sigma models (NLSM) coupled to nontrivial DCFTs.  
\item We provide an explicit example of a unitary DQFT that exhibits SSB. The example is particularly simple: it consists of $N$ free massless scalars in $4<d<6$ spacetime dimension coupled to a $O(N)/O(N-1)$ nonlinear sigma model on a two-dimensional surface.  When the bulk field is integrated out, the model becomes similar to the two-dimensional long-range $O(N)$ model. In agreement with the first result, the defect RG flow never reaches a fixed-point in the infrared as a consequence of the moduli space of the massless scalars. 
\item By studying NLSMs on defects coupled to nontrivial DCFTs we find a generic kind of defect universality class, in which there is no SSB and correlation functions decay logarithmically with the distance.  This can be understood as a DCFT perturbed by a marginally irrelevant interaction term and it is analogous to the recently found extraordinary-log surface universality class in the $3d$ $O(N)$ model \cite{Metlitski:2020cqy,Krishnan:2023cff}.  The analysis of such phase in general depends on the geometry of the coset describing the symmetry breaking pattern and some model-dependent nonperturbative DCFT data.
\end{itemize}
We expect that our results rule out the possibility of SSB of continuous internal symmetries in most examples of interest. Additionally, our EFT analysis should describe a large class of defect RG flows on surface defects.  We detail below the organization of the rest of the paper and the content of its section.

In sec.~\ref{sec_Coleman_review} we review Coleman's theorem for two-dimensional QFTs.  Besides being a useful warm-up for the analysis of surface defects, this section also serves the (modest) goal of providing a modern and relatively complete exposition to Coleman's theorem. To this aim, we first discuss in detail a nonperturbative argument in the spirit of Coleman \cite{Coleman:1973aa} in sec.~\ref{subsec_CMW_thm}. We then exemplify the theorem with EFT methods, by studying generic NLSMs for the Goldstone modes in sec.~\ref{subsec_CMW_NLSMs}.  The EFT analysis also provides explicit conditions for the absence of SSB in nonunitary two-dimensional models,  that may be of interest in statistical mechanics, for which the general Coleman's theorem does not apply. We finally discuss explicit examples in sec.~\ref{subsec_CMW_ex}.

In sec.~\ref{sec_SSB_WI} we discuss SSB on surface defects at the non-perturbative level. We first review the structure Ward identities for internal symmetries in DQFTs in sec.~\ref{subsec_DQFT_review}. We then use Ward identities to analyze how Coleman's argument is modified when considering surface defects rather than local QFTs in sec.~\ref{subsec_SSB_defects}. Our analysis highlights how Coleman's (and Goldstone's) arguments may fail for DQFTs. Finally we show that SSB is inconsistent when the theory flows to a scale invariant fixed point.

In sec.~\ref{sec_free_ex} we discuss the example of a free scalar in $d>4$ spacetime dimensions, coupled to a $O(N)$ nonlinear sigma model on a surface defect.   We show that SSB occurs for $d<6$, where the RG flow is of runaway type, but not for $d\geq 6$, in which case the model flows to a DCFT at large distances. Our results agree with (and extend) the known constraints on SSB in the long-range $O(N)$ model \cite{Mermin:1966fe,Halperin},  providing a simple field-theoretical perspective based on defect RG.

In sec.~\ref{sec_log} we discuss DCFTs coupled to a NLSM on a $p$-dimensional defect. As we argue in sec.~\ref{sec_SSB_WI}, this setup is expected to describe the generic endpoint of an RG flow for a DQFT that breaks spontaneously a continuous internal symmetry. After presenting the setup in sec.~\ref{subsec_log_setup}, in sec.~\ref{subsec_log_RG} we show that quantum effects ensure that no SSB occurs when we extrapolate such a low energy EFT to surface defects. Interestingly, despite the absence of SSB, sometimes the NLSM remains weakly coupled at arbitrary large distances (also for non-Abelian groups). In this case, the endpoint of the RG flow can be understood as a DCFT perturbed by a marginally irrelevant interaction. We show that this leads to correlation functions that decay logarithmically. Our analysis generalizes the finding of \cite{Metlitski:2020cqy} for boundaries in the $O(N)$ models to arbitrary symmetry breaking patterns. We discuss that and other examples in sec.~\ref{subsec_log_ex}.

\section{Coleman's theorem in \texorpdfstring{$2d$}{2d} QFT}\label{sec_Coleman_review}

\subsection{General argument}\label{subsec_CMW_thm}

Let us specify the setup and our assumptions. We consider a QFT in two spacetime dimensions. The theory is invariant under an internal continuous $0$-form symmetry group $G$, that is a compact
Lie group generated by the charges $\{Q_a,Q_i\}$.  Our main assumption is that the vacuum is invariant under the Poincar\'e group and it admits a unitary Hilbert space when quantized in Lorentzian signature. This in particular allows expressing all two-point functions (both in Euclidean and Lorentzian signature) in terms of the spectral density.\footnote{More precisely, this is true up to contact terms. These are irrelevant at finite separation and thus do not play a role in our discussion.}

The argument proceeds by contradiction.  Suppose that the ground-state breaks the internal symmetry group to a (possibly trivial) subgroup $H$,  generated by the charges $\{Q_i\}$.  This means that there is at least a local operator which transforms in a (not necessarily irreducible) representation $R$ of $G$ whose expectation value is not invariant under the action of the internal symmetry group\footnote{We work in Euclidean signature in standard conventions such that the charges are anti-Hermitian.}
\begin{equation}\label{eq_SSB}
\langle \mO_A\rangle= v_A\quad\text{such that}\quad \langle [Q_a,\mO_A]\rangle \equiv -\delta_a v_A\neq 0\,,
\end{equation}
where the set of the $\{Q_a\}$ furnishes a basis of generators for the coset $G/H$.  We will show that eq. \eqref{eq_SSB} is incompatible with both reflection positivity and the cluster decomposition principle.

As well known, eq. \eqref{eq_SSB} implies the existence of $n_G-n_H$ massless particles, the Goldstone bosons, where $n_G$ and $n_H$ are, respectively, the dimensions of $G$ and $H$.  Importantly,  Goldstone bosons have non-zero matrix element with the order parameter.  For completeness and in view of the future application to defects, we review the proof of these claims  below.

Consider the Euclidean two-point function of the current and the order parameter
\begin{equation}
\langle J^\mu_a(x) \mO_A(0)\rangle=\int\frac{d^2p}{(2\pi)^2}e^{-i p x}H^\mu(p)\,.
\end{equation}
Eq. \eqref{eq_SSB} is equivalent to the following Ward identity
\begin{equation}\label{eq_WI_bulk}
\langle \pd_\mu J^\mu_a(x) \mO_A(0)\rangle=-\delta_a v_A\delta^2(x)\,,
\end{equation}
from which we conclude that the correlator in momentum space reads
\begin{equation}\label{eq_Hmu}
 H^\mu(p)=-i\delta_a v_A\frac{p^\mu}{p^2}+ i\varepsilon^{\mu\nu}p_\nu f(p^2)\,,
\end{equation}
for an arbitrary function $f(p^2)$. (In parity invariant theories $f(p^2)=0$.) The pole for $p^2\rightarrow 0$ implies the existence of a delta function contribution in the spectral density of the correlator, as it follows from the K\"allen-Lehmann decomposition 
\begin{equation}\label{eq_H_rho}
H^\mu(p)= \int_0^{\infty} dm^2 \frac{p^\mu \rho_P(m^2)+\varepsilon^{\mu\nu}p_\nu \rho_A(m^2)}{p^2+m^2}\,,
\end{equation}
where, in obvious notation,\footnote{Here and in the following, the Wick rotation $V^0\rightarrow i V^0$ is understood for all vectors $V^\mu$ in Lorentzian matrix elements.}
\begin{equation}\label{eq_H_rho2}
p^\mu \rho_P(p^2)+\varepsilon^{\mu\nu}p_\nu \rho_A(p^2)=
(2\pi)\sum_n\langle 0|J_a^\mu(0)|n\rangle\langle n|\mO_A(0)|0\rangle\delta^2(p-q_n)\,.
\end{equation}

Eq. \eqref{eq_Hmu} hence implies $\rho_P(m^2)=-i\delta_a v_A\delta(m^2)$ exactly. As well known, a delta function in the spectral density of local operators cannot arise from the continuum part of the spectrum. Indeed this would imply a singular matrix element $\propto \delta(m^2)$ between either the current or the order paramater and a state belonging to the continuum part of the spectrum. However,  by application of the spectral decomposition, such a singular matrix element would imply that the two-point function of this local operator with itself diverges at any finite distance. Notice that it is important that we are considering local operators for this argument; this observation will be important when we will analyze the consequences of a similar argument for defects.

We therefore conclude that there exist $n_G-n_H$ massless particles $\pi^a$. Importantly, these are interpolated both by the current and by the order parameter, with matrix elements given by 
\begin{equation}\label{eq_matrix_elements}
\langle 0| J_a^\mu(0)|\pi^b(p)\rangle= ip^\mu \frac{f_{ab}}{\sqrt{4\pi p^0}}+i\varepsilon^{\mu\nu}p_\nu \frac{f^{(A)}_{ab}}{\sqrt{4\pi p^0}}\,,\qquad
\langle \pi^b(p)|\mO_A(0)|0 \rangle= \frac{Z^b_{A}}{\sqrt{4\pi  p^0}}\,,
\end{equation}
such that $f_{ab} Z^b_{A}=-\delta_a v_A$ and $f^{(A)}_{ab}$ is not determined by Goldstone's theorem.

Eq. \eqref{eq_matrix_elements} implies the existence of a delta function in the spectral density of the order parameter connected two-point function
\begin{equation}\label{eq_rho_AB}
\begin{split}
\rho_{AB}(p^2)&=
(2\pi)\sum_{n\neq 0}\langle 0|\mO_A^\dagger(0)|n\rangle\langle n|\mO_B(0)|0\rangle\delta^2(p-q_n) \\
&=M_{AB}^{\pi}\delta(p^2)+\delta\rho_{AB}(p^2)\,,
\end{split}
\end{equation}
where we isolated the contribution of the Goldstone bosons, given by the matrix $M_{AB}^\pi=\sum_a(Z_{A}^{a})^*Z^a_{B}$. Unitarity implies that both contributions are semi-positive definite
\begin{equation}
M_{AB}^\pi\succeq 0 \,,\qquad
\delta\rho_{AB}(m^2)\succeq 0\,.
\end{equation}

To show the contradiction we announced at the beginning, let us look at the first term in the second line of equation \eqref{eq_rho_AB}. It tells us that the order parameter two-point functions become very negative when the distance is very large. This behaviour violates both the cluster decomposition principle and reflection positivity.\footnote{This also implies that Fourier-transform of the correlator $\langle \mO_A^\dagger(x)\mO_B(0)\rangle_c$ is not well-defined due to infrared divergences. Often, this infrared divergence is regarded as an inconsistency of the theory \emph{per se}. While this is ultimately correct, in our argument we want to highlight the role of the physical principles of reflection positivity and clustering.} This happens because the massless particle propagator is logarithmic at large distances:
\begin{equation}\label{eq_log}
\langle \mO_A^\dagger(x)\mO_B(0)\rangle_c\sim -\log( |x|)M^{\pi}_{AB}\quad
\text{for }|x|\rightarrow\infty\,.
\end{equation}
To prove this last statement at the non-perturbative level, it is convenient to consider the derivative of the two-point function at distance $|x|=r$
\begin{equation}
F_{AB}(r)=-
2\pi r\pd_r\langle\mO_A(x)\mO_B(0)\rangle_c\vert_{|x|=r}\,.
\end{equation}
This object admits a K\"allen-Lehman decomposition in the form\footnote{Notice that eq. \eqref{eq_F} holds also when $\delta\rho_{AB}$ contains delta functions $\delta(m^2)$,  since $\lim_{z\rightarrow 0}z K_1(z)=1$.}
\begin{equation}\label{eq_F}
\begin{split}
F_{AB}(r) &=-2\pi
\int_0^\infty dm^2\rho_{AB}(m^2)\int \frac{d^2p}{(2\pi)^2}\frac{ i p x \,e^{ipx}}{p^2+m^2}
\\& =M^{\pi}_{AB}+\int_0^\infty dm^2 \delta\rho_{AB}(m^2)rm  K_1( r m)\,,
\end{split}
\end{equation}
where $K_1(z)$ is a modified Bessel function of the second kind. Notice that, while the Fourier transform of the massless propagator is famously infared divergent, there are no ambiguities of this sort in eq.~\eqref{eq_F} thanks to the derivative. Since $K_1(z)>0$ for $z>0$,  both terms on the right hand-side of eq.~\eqref{eq_F} are semi-positive definite and we conclude that
\begin{equation}
F_{AB}(r)=-2\pi r\pd_r \langle\mO_A(x)\mO_B(0)\rangle_c\vert_{|x|=r}\succeq M_{AB}^{\pi}\,.
\end{equation}
By contracting this equation with two nontrivial eigenvectors of $M_{AB}^\pi$ we find a strict inequality.  This implies that at least some components of the correlation function will keep decreasing to arbitrary negative values as the distance $r$ increases,  violating hence both the cluster decomposition principle and reflection positivity as announced.  We conclude therefore that eq.~\eqref{eq_SSB} is inconsistent in a local unitary $2d$ QFT.

Some comments are in order
\begin{itemize}
\item Physically, the absence of symmetry breaking is due to the fact that massless excitations in $2d$ propagate to arbitrary large distances, cfr. eq.~\eqref{eq_log}. This means that the ordered phase is unstable to arbitrary small local perturbations. Symmetry breaking is thus impossible in quantum (and statistical) theories due to vacuum fluctuations.\footnote{For Abelian symmetries, this argument can be made concrete by considering the time evolution of a wave-functional which is localized in field space at some initial time. One finds that the wave-functional spreads out, and at very late times the state is totally delocalized \cite{Endlich:2010hf}. Equivalently, one can prove that the energy cost of a domain wall is independent of the size of the system \cite{Imry:1975zz}. }
\item The two main ingredients of the theorem are Poincar\'e invariance, which constrains the structure of the spectral decomposition \eqref{eq_H_rho}, and unitarity, which implies that both terms in eq. \eqref{eq_F} are semi-positive definite and at least one of them is nonzero.  It is indeed possible to construct examples where spontaneous symmetry breaking (SSB) occurs in two-dimensional theories which violate either of these assumptions.\footnote{In theories without Lorentz invariance (or for states in which the latter is nonlinearly realized as in the presence of a chemical potential), it is well known that it is possible to have type II Goldstones, i.e. with dispersion relation $\omega\sim \vec{k}^{\,2}$, in one spatial dimension \cite{Brauner:2010wm,Watanabe:2019xul}. An example of a relativistic nonunitrary theory which exhibits SSB, 
is the $O(N-1,1)/O(N-1)$ noncompact nonlinear sigma model for $N\geq 2$, which is weakly coupled at low energies and violates the condition \eqref{eq_gamma_pos} - see the next subsection. 
Another example is provided by two free compact bosons with Lagrangian $\mL=\frac{R}{2}[(\pd\phi_1)^2-(\pd\phi_2)^2]$, for which the operators $\mO_{q,\pm}=\exp\left[ i q(\phi_1\pm\phi_2)\right]$ (with $q\in\mathds{Z}$) act as order parameters breaking the $U(1)\times U(1)$ symmetry. 
Additional physically interesting violations of Coleman's theorem are described in \cite{Halperin}.}
\item Coleman's theorem does not necessarily imply the absence of massless particles in two-dimensions. It is crucial that these particles have a non-zero matrix element with a scalar operator, as in eq.~\eqref{eq_matrix_elements}, to reach a contradiction.  In other words, Coleman's theorem only forbids order parameters. Indeed, the free boson is an example of a two-dimensional theory with a massless particle.
\item As already remarked below eq. \eqref{eq_H_rho2}, it is crucial that the Ward identity almost fully determines the two-point between two \emph{local} operators in order to prove Goldstone's theorem. For defects, the Ward identities are more involved and we will see that Goldstone's theorem does not always hold.
\item A potentially confusing aspect of the Coleman's argument is that, despite the positivity of the spectral density, one finds that eq.~\eqref{eq_SSB} implies that correlation functions of the order parameter violate relfection positivity. This seems in contradiction with the Osterwalder-Schrader reconstruction theorem \cite{Osterwalder:1973dx,glimm2012quantum}, which states the equivalence between Euclidean reflection positivity and Lorentzian unitarity in QFT. The resolution of the paradox is that the Osterwalder-Schrader reconstruction theorem assumes also the cluster decomposition principle, which is explicitly violated by the 2pt. function~\eqref{eq_log}.
\end{itemize}

\subsection{Nonlinear sigma models}\label{subsec_CMW_NLSMs}

\subsubsection{The action}

To understand how the theorem is realized in examples, it is instructive to consider the most general low energy effective-field theory for the would-be-Goldstone fields.  This is the nonlinear sigma model (NLSM) with target space given by the homogeneous manifold $G/H$.

We denote the broken generators $Q_a$ and the unbroken ones $Q_i$ as in the previous section.\footnote{With a slight abuse of notation, we denote the abstract group generators and the corresponding QFT operators with the same symbols.} The algebra is parametrized by the fully antisymmetric structure constants as
\begin{equation}\label{eq_algebra}
[Q_a,Q_b]=if_{abc} Q_c+i f_{abi}Q_i\,,\qquad
[Q_a,Q_i]=i f_{aib} Q_b\,,\qquad
[Q_i,Q_j]=i f_{ijk} Q_k\,,
\end{equation}
The Goldstone fields $\pi^a$ span the coset manifold $G/H$. A convenient parametrization for an arbitrary element $\Omega\in G/H$ close to the identity is
\begin{equation}\label{eq_coset}
\Omega=\exp\left[i \pi^a(x) Q_a\right]\,.
\end{equation}
The group $G$ acts on the left of the coset $\Omega$. This means that the Goldstone fields transform as  $\pi\rightarrow \pi'$, where
\begin{equation}\label{eq_transformation_rule}
g\Omega(\pi)=\Omega(\pi')h(\pi')\,,\quad
h(\pi')\in H\,.
\end{equation}
In particular,  a transformation $g=\exp ( i\alpha^a Q_a)$ acts as $\pi^a\rightarrow\pi^a+\alpha^a$ to linear order in the fields and the transformation parameter.

To build the most general action we follow the CCWZ construction \cite{Coleman:1969sm,Callan:1969sn} and consider the Maurer-Cartan one-form:
\begin{equation}
\begin{split}
\Omega^{-1}d \Omega
&=i \left(D_\mu\pi^a Q_a+ A^i_\mu Q_i\right)dx^\mu\,.
\end{split}
\end{equation}
Under the action of the symmetry group $G$ the \emph{covariant derivatives} $D_\mu\pi^a$ transform via a linear (local) action of $H$: $D_\mu\pi^a Q_a\rightarrow D_\mu\pi^a h^{-1}Q_a h$, where $h$ is defined in eq. \eqref{eq_transformation_rule}.  The $A^i_\mu$ transform as connections, similarly to gauge fields. The explicit expressions for the covariant derivatives and the connections can be found as a series expansion in the fields using the algebra \eqref{eq_algebra}:
\begin{align} \nonumber
D_\mu\pi^a 
&=\pd_\mu\pi^a-\frac12f_{bac}\pi^b\pd_\mu\pi^c +\frac16
(f_{bae}f_{ced}+f_{bai}f_{cid})\pi^b\pi^c\pd_\mu\pi^d +O\left(\pi^3\pd\pi\right)\,,\\
A_\mu^i 
&=-\frac12 f_{aib}\pi^a\pd_\mu\pi^b 
+\frac16 (f_{a i d} f_{b d c}+ f_{a i j} f_{b j c}) \pi^a\pi^b\pd_\mu\pi^c +O\left(\pi^3\pd\pi\right)\,. \label{eq_D_A_CCWZ}
\end{align}

It is now simple to construct the most general action for the Goldstone fields to leading order in derivatives.  We restrict to parity invariant actions. We simply use the covariant derivatives $D_\mu\pi^a$ as building blocks, appropriately contracted via an $H$-invariant tensor that we denote $g _{ab}(0)$: 
\begin{equation}\label{eq_NLSM}
\mL=\frac{1}{2}D_\mu \pi^a D_\mu \pi^b g _{ab}(0)=
\frac12 \pd_\mu\pi^a\pd_\mu\pi^b g_{ab}(\pi)\,,
\end{equation}
where in the last equality we defined the metric $g_{ab}(\pi)$ of the NLSM.  Unitarity demands that the tensor is $g _{ab}(0)$ positive-definite and symmetric.  The connection $A^i_\mu$ may further be used to build higher derivative corrections.\footnote{In the absence of parity one may, in some cases, construct additional terms, such as $D_\mu\pi^a D_\nu\pi^b\varepsilon^{\mu\nu} f_{ab}$, for some antisymmetric tensor $f_{ab}$ invariant under $H$ (one such term is the 2d theta-angle $n_A\pd_\mu n_B \pd_\nu n_C\varepsilon^{ABC}\varepsilon^{\mu\nu}$ in the $O(3)/O(2)$ NLSM, see sec.~\ref{subsec_CMW_ex} for the notation). Sometimes one can also construct generalized Wess-Zumino terms out of the $A^i_\mu$. The latter have quantized coefficients and, in perturbation theory, we can neglect them as long as the coupling $g ^{ab}(0)$ is much smaller than the inverse of all the Wess-Zumino coefficients. It is also known that in general Wess-Zumino terms (and sometimes the generalized theta angles) make the theory flow to a $CFT_2$ at finite coupling where the symmetry is linearly realized. } For simplicity, we neglect the possibility of having massless fields not arising from SSB of an internal symmetry.

Finally, it is also convenient to supplement the action \eqref{eq_NLSM} with a mass term which breaks explicitly the symmetry $G$ and acts as an infrared regulator. We take it to be
\begin{equation}\label{eq_L_IR}
\mL_{IR}\simeq\frac12 m_{IR}^2 \pi^a  \pi^bg _{ab}(0)\,.
\end{equation}
This choice is particularly convenient since it preserves the unbroken group $H$ and leads to a simple expression for the $2d$ propagator of the Goldstone fields
\begin{equation}\label{eq_prop}
\langle\pi^a(x)\pi^b(0)\rangle=g^{ab}(0)\int \frac{d^2k}{(2\pi)^2}\frac{e^{ikx}}{k^2+m_{IR}^2}
\equiv g^{ab}(0)G(x)\,.
\end{equation}

Notice that the chosen regulator~\eqref{eq_L_IR} is incompatible with the periodicity of the Goldstone fields, but this will not be an issue for our perturbative analysis; the mass term~\eqref{eq_L_IR} provides a simple infrared cutoff compatible with all spacetime symmetries in loop calculations.  One could in principle also introduce an infrared regulator at the full nonperturbative level by supplementing the term \eqref{eq_L_IR} with nonlinear corrections, so to preserve the periodicity of the Goldstone fields.  As we will only work in perturbation theory, we neglect these corrections in what follows.\footnote{More precisely, the nonlinearities needed to preserve the periodicity of the Goldstones can be neglected only when computing $n\geq 2$ point functions of operators at distances $|x_{ij}|\ll m_{IR}^{-1}$. Therefore, to compute one-point functions, such as the expectation value of the order parameter, we should consider the $2-$point function $\langle\mO_A^\dagger(x)\mO_B(0)\rangle$ in the limit $|x|\rightarrow \infty$, $m_{IR}\rightarrow 0$ with $|x|m_{IR}\rightarrow 0$.  In practice the same result is obtained by direct calculation of the one-point function assuming that the infrared regulator is given \emph{exactly} by eq.~\eqref{eq_L_IR} and taking the limit $m_{IR}\rightarrow 0$. }

\subsubsection{Perturbative RG flows}

In dimensions $d>2$ the coupling of the NLSM \eqref{eq_NLSM}, which is given by the inverse metric at the origin $g^{ab}(0)$,  is irrelevant. Thus quantum effects do not play an important role and SSB surivives in the quantum theory. In two dimensions instead the coupling is marginal, and the fate of SSB at a quantum level depends on the RG flow of this coupling. In this section we review some basic aspects of this RG flow.

As well known from string theory,  the beta function of a NLSM can be written in terms of the Riemann tensor derived from $g_{ab}(\pi)$ \cite{Friedan:1980jm,Friedan:1980jf,ketov2000quantum}:  
\begin{equation}\label{eq_beta_NLSM}
\beta_{ab}=\frac{\pd g_{ab}}{\pd\log\mu}=
\frac{1}{2\pi}\mathcal{R}_{ab}+\frac{1}{8\pi^2} \mathcal{R}^2_{ab}+O\left(\mathcal{R}^3\right)\,,
\end{equation}
where $\mu$ denotes the RG scale, $\mathcal{R}_{ab}$ is the Ricci tensor,  and $\mathcal{R}_{ab}^2=\mathcal{R}^c_{\;ade}\mathcal{R}^d_{\;bfm}g_{cd}g^{df}g^{em}$. From here on all geometric invariants are evaluated at $\pi^a=0$ unless specified otherwise.

The loop expansion is equivalent to an expansion in derivatives of the metric, the $n$th order contributing schematically as $\sim \mathcal{R}^n$.  The perturbative regime thus corresponds to small curvature. A general analysis of RG flows in NLSMs was given in \cite{Friedan:1980jm,Friedan:1980jf}.  It is known that NLSMs with target space given by a homogeneous manifold are always asymptotically free. In particular, the theory remains perturbative at arbitrary scales in the infrared if and only if the curvature of the coset manifold $\mathcal{R}^a_{\;bcd}$ is trivial, i.e. if $G$ and $H$ are a product of Abelian factors $U(1)\times U(1)\ldots \times U(1)$.  

In view of the future generalization to the defect case, let us sketch the analysis which leads to this result. Geometrically, the beta-function \eqref{eq_beta_NLSM} describes a flow in the space of $G$ invariant metrics on the manifold $\mathcal{M}=G/H$; this is fully specified by the $H$-invariant tensor $g_{ab}(0)$. The fixed-points are the solutions to the equation $\beta_{ab}=0$.\footnote{More in general,  eq.~\eqref{eq_beta_NLSM} transforms covariantly under field-redefinitions corresponding to a diffeormorphism preserving the isometry group $G$. Therefore the fixed-points of the beta-function \eqref{eq_beta_NLSM} are the metrics $g_{ab}$ for which $\beta_{ab}$ is tangent to the orbit of the diffeomorphism group \cite{Friedan:1980jm}:
\begin{equation}\label{eq_beta_v}
\beta_{ab}=\nabla_a v_b+\nabla_b v_a\,,
\end{equation}
where $v$ is a left-invariant vector under $G$.  When $v_a=\pd F /\pd\pi^a$ for some function $F$ of the fields, the fixed point is conformally invariant and we can perform a scheme redefinition after which $\beta_{ab}=0$ \cite{Nakayama:2013is}; see e.g. for the discussion of similar ambiguities in beta functions \cite{Luty:2012ww,Fortin:2012hn,Baume:2014rla}. When $v_a$ is not the gradient of a scalar, the fixed point is only scale invariant. In unitary two-dimensional theories in $2d$, it is well know that scale invariance implies conformal invariance \cite{Polchinski:1987dy} and thus we neglect this possibility.  See \cite{Nakayama:2020bsx} for a more extensive discussion of scale and conformal invariance in nonlinear sigma models. We thank Yu Nakayama for useful correspondence on this point. \label{foot_beta_v}} We follow \cite{Friedan:1980jm} and decompose the metric as $g_{ab}=\hat{g}_{ab}/T$, where the hatted metric is normalized so that the volume element $\sqrt{\det(\hat{g})}$ is constant along the RG.  Thus, $T$ roughly characterizes the coupling strength. Setting $t=-\log(\mu/\mu_0)$, we find the following system of equations: 
\begin{align}\label{eq_RG_T}
\frac{\pd T}{\pd t}&=\frac{T^2}{2\pi n}\hat{\mathcal{R}}+\frac{T^3}{8\pi^2n}\hat{\mathcal{R}}_{ab}^2\hat{g}^{ab}+O\left(T^4\right)\,,\\
\label{eq_RG_g}
\frac{\pd \hat{g}_{ab}}{\pd t}&=\frac{T}{2\pi}\left(\hat{\mathcal{R}}_{ab}-
\frac1n\hat{g}_{ab}\hat{\mathcal{R}}\right)+\frac{T^2}{8\pi^2}\left(\hat{\mathcal{R}}^2_{ab}-
\frac{1}{n}\hat{g}_{ab}\hat{\mathcal{R}}_{cd}^2\hat{g}^{cd}\right)+O\left(T^3\right)\,,
\end{align}
where $n=n_G-n_H$ is the dimension of the manifold.

Naively, one might think that $T=0$ with arbitrary $\hat{g}_{ab}$ provides a fixed-point. However, since in parturbative RG flows hatted curvature invariants $\sim\hat{\mathcal{R}}$ remain bounded,  eq.~\eqref{eq_RG_T} implies that $T$ approaches the point $T=0$ as $1/t$ or slower for $t\rightarrow \pm\infty$, where the sign depends on whether the fixed-point is an IR or a UV one. Therefore we can change variable as $t\rightarrow y=\int_{t_0}^t dt' \,T(t')$ in eq.~\eqref{eq_RG_g}, such that $y\rightarrow\pm\infty$ for $t\rightarrow\pm\infty$.  We then find that also the following condition needs to be satisfied
\begin{equation}\label{eq_RG_Einstein}
\hat{\mathcal{R}}_{ab}=\frac1n\hat{g}_{ab}\hat{\mathcal{R}}\,.
\end{equation}
Eq.~\eqref{eq_RG_Einstein} defines the so-called \emph{Einstein manifolds}. A general result in differential geometry is the following: all compact homogeneous Einstein manifolds have nonnegative Ricci scalar $\hat{\mathcal{R}}\geq 0$, with the case $\hat{\mathcal{R}}=0$ trivially associated with a flat torus, i.e. a fully Abelian group \cite{jablonski2021survey}. Using this result in eq.~\eqref{eq_RG_T} we immediately find that fixed-points $T=0$ with $\hat{g}_{ab}$ satisfying eq.~\eqref{eq_RG_Einstein}, when they exist, are necessarily unstable, unless the group is Abelian, in which case $T$ is marginal.  In the non-Abelian case, the coupling $T$ always increases to non-perturbative values.

We can analyze correlation functions reliably as long as the coupling remains perturbative. Let us consider in particular the order parameter $\mO_A$ which transforms in a (not necessarily irreducible) representation $R$ of $G$. In general, the order parameter corresponds to multiple independent local operators. In the low energy EFT we represent $\mO_A$ matching its quantum numbers in terms of the Goldstone degrees of freedom.  To this aim, we introduce a constant vector $v_A$ which is left invariant by the unbroken group $H$:
\begin{equation}\label{eq_v}
v_A=\{v_{\bar{A}},\vec{0}\}\,\quad
\text{s.t. }
(Q_i)^{(R)}_{AB}v_B=0\,.
\end{equation}
In eq.~\eqref{eq_v}, we denoted with a bar the components upon which the unbroken generators act trivially, i.e.  $(Q_i)^{(R)}_{A\bar{B}}=0$. 
The value of the nontrivial entries $v_{\bar{A}}$ is arbitrary within EFT.
The order parameter is then given by
\begin{equation}\label{eq_order_parameter}
\mO_A=\left[\Omega(\pi)\right]^{(R)}_{AB} v_B\,,
\end{equation}
where $(Q)^{(R)}$ and $\left[\Omega(\pi)\right]^{(R)}$ denote, respectively,  the appropriate representation of the group generators and the coset \eqref{eq_coset}. The number of independent operators corresponding to eq.~\eqref{eq_order_parameter} coincides with the number of independent constants $v_{\bar{A}}$ in eq.~\eqref{eq_v}.   At higher order in derivatives the invariants in eq.~\eqref{eq_D_A_CCWZ} can be used to include additional corrections to the operator matching~\eqref{eq_order_parameter}; these corrections do not contribute to the one-point function of the operator and are thus negligible for our purposes.

At a classical level, the expectation value of the order parameter is given by
\begin{equation}\label{eq_O_class}
\langle\mO_{\bar{A}}\rangle= v_{\bar{A}}\,.
\end{equation}
It is interesting to analyze how this result changes upon including quantum corrections. To this aim we expand eq.~\eqref{eq_O_class} in the fields
\begin{equation}\label{eq_O_quant_pre}
\langle\mO_A(0)\rangle=v_A+i(Q_a)^{(R)}_{AB}v_B\langle\pi^a(0)\rangle-\frac{1}{2}\left(Q_a Q_b\right)^{(R)}_{AB}v_B\langle\pi^a(0)\pi^b(0)\rangle+\ldots\,.
\end{equation}
The term $\langle\pi^a(0)\rangle$ seemingly receives a contribution at one-loop from the cubic vertices in the action \eqref{eq_NLSM}. It may however  be checked that this contribution vanishes. The self-contraction is evaluated using the propagator \eqref{eq_prop}:
\begin{equation}
\langle\pi^a(0)\pi^b(0)\rangle=
\int \frac{d^2p}{(2\pi)^2}\frac{g^{ab}}{p^2+m_{IR}^2}
=\frac{g^{ab}}{2\pi}\log\left(\Lambda/m_{IR}\right)\,,
\end{equation}
where $\Lambda$ is the cutoff scale.  We remove the dependence on the cutoff scale $\Lambda$ introducing the wave-function renormalization as $v_{\bar{A}}\rightarrow Z_{\bar{A}\bar{B}}v_{\bar{B}}$. We choose
\begin{equation}\label{eq_Z}
Z_{\bar{A}\bar{B}}=\delta_{\bar{A}\bar{B}}-\frac{1}{4\pi}g^{ab}
\left(Q_a Q_b\right)^{(R)}_{\bar{A}\bar{B}}\log\left(\Lambda/\mu\right)\,,
\end{equation}
where $\mu$ is the sliding scale.  We thus find that the expectation value of the renormalized operator reads
\begin{equation}\label{eq_O_quant_ren}
\langle\mO_{\bar{A}}\rangle\simeq v_{\bar{A}}-\gamma_{\bar{A}\bar{B}}v_{\bar{B}}\log\left(\mu/m_{IR}\right)
\equiv v_{\bar{A}}^{(r)}(\mu/m_{IR},g^{ab}(\mu))
\,,
\end{equation}
where $g^{ab}(\mu)$ denotes the coupling evaluated at the scale $\mu$ and $\gamma_{\bar{A}\bar{B}}$ is the anomalous dimension matrix, defined by
\begin{equation}\label{eq_anomalous}
\gamma_{\bar{A}\bar{B}}=\left(Z^{-1}\frac{\pd Z}{\pd\log\mu}\right)_{\bar{A}\bar{B}}=
\frac{1}{4\pi}g^{ab}
\left(Q_a Q_b\right)^{(R)}_{\bar{A}\bar{B}}\,.
\end{equation}
Importantly, for a compact group the matrices $Q_a^{(R)}$ are Hermitian and thus the anomalous dimension matrix is strictly positive definite
\begin{equation}\label{eq_gamma_pos}
\gamma_{\bar{A}\bar{B}}\succ 0\,.
\end{equation}

In summary, we see that the one-loop correction to the expectation value diverges for $m_{IR}\rightarrow 0$, signalling the breakdown of the classical approximation.  Physically, as we lower the IR regulator $m_{IR}$, quantum effects become important and erase the classical expectation value. This claim can be made precise using the Callan-Symanzik equation
\begin{equation}
\left(\mu\frac{\pd}{\pd\mu}+\beta_{ab}\frac{\pd}{\pd g_{ab}}\right) v^{(r)}_{\bar{A}}+\gamma_{\bar{A}\bar{B}} v^{(r)}_{\bar{B}}=0\,,
\end{equation}
whose general solution reads
\begin{equation}\label{eq_sol_CS}
v_{\bar{A}}^{(r)}(\mu/m_{IR},g^{ab}(\mu))=\left[\exp\left(-\int^{\log\mu}_{\log m_{IR}} d\log\bar{\mu}\,\gamma(g^{ab}(\bar{\mu}))\right)\right]_{\bar{A}\bar{B}}
v_{\bar{B}}^{(r)}(1,g^{ab}(m_{IR}))\,.
\end{equation}
The meaning of this equation is particularly clear in a regime of the RG flow where the running of the coupling can be neglected - as it happens for Abelian groups. In this case, the anomalous dimension matrix does not depend on $\mu$ and we can decompose the classical expectation value $v_{\bar{A}}$ into eigenvectors of $\gamma$:
\begin{equation}
v_{\bar{A}}=\sum_{\alpha} c_{\alpha} w^{(\alpha)}_{\bar{A}}\quad
\text{such that}\quad\
\gamma_{\bar{A}\bar{B}}w^{(\alpha)}_{\bar{B}}=\bar{\gamma}_\alpha w^{(\alpha)}_{\bar{A}}\,,
\end{equation}
where $\bar{\gamma}_\alpha>0$.  Neglecting the dependence on the coupling, eq.~\eqref{eq_sol_CS} then reads
\begin{equation}\label{eq_O_CS_bulk}
\langle \mO_{\bar{A}}\rangle\approx \sum_{\alpha} c_{\alpha} w^{(\alpha)}_{\bar{A}}\left(\frac{m_{IR}}{\mu}\right)^{\bar{\gamma}_{\alpha}}\,.
\end{equation}
Since the anomalous dimension matrix is positive definite, eq.~ \eqref{eq_O_CS_bulk} shows that the one-point function of the order parameter approaches zero for $m_{IR}\rightarrow 0$. The result \eqref{eq_O_CS_bulk} is exact for Abelian groups, while in the non-Abelian case the dependence on the coupling constant becomes important as we decrease $m_{IR}$. In that case, the NLSM becomes strongly coupled at a scale $m_{IR}\sim \Lambda_{IR}$. At that scale the Goldstone modes are generically expected to develop a gap and the symmetry is restored.

We remark that the expressions for the anomalous dimensions \eqref{eq_anomalous} and the solution to the Callan-Symanzik equation \eqref{eq_sol_CS} do not depend in any way on the asymptotic freedom of the NLSM.  In fact,  the structure of perturbative RG flows may change in models with additional light degrees of freedom. Additionally, some non-unitary NLSMs which are of interest in statistical mechanics are not asymptotically free and remain perturbative in the infrared. It is thus meaningful to ask under which conditions on the RG flow of the coupling $g^{ab}$ do these expressions imply that the expectation value of the order parameter vanishes for $m_{IR}\rightarrow 0$:
\begin{equation}\label{eq_no_SSB_cond}
\langle \mO_{\bar{A}}\rangle=v_{\bar{A}}^{(r)}(\mu/m_{IR},g^{ab}(\mu))
\xrightarrow{m_{IR}\rightarrow 0} 0\,.
\end{equation} 
Assuming a fully perturbative RG flow, so that the only relevant dependence on the coupling is through the anomalous dimensions,  eq. \eqref{eq_no_SSB_cond} is equivalent to
\begin{equation}\label{eq_int_condition}
\lim_{m_{IR}\rightarrow 0^+}\left[\exp\left(-\int^{\log\mu}_{\log m_{IR}} d\log\bar{\mu}\,\gamma(g^{ab}(\bar{\mu}))\right)\right]_{\bar{A}\bar{B}} v_{\bar{B}}=0\,.
\end{equation}
A moment of reflection shows that a sufficient condition for this to happen for arbitrary values of $v_{\bar{A}}$ is given by
\begin{equation}\label{eq_g_condition}
\lim_{\mu\rightarrow 0^+} (-\log\mu)g^{ab}(\mu)\succ 0\,.
\end{equation}
When the condition \eqref{eq_g_condition} is saturated $g^{ab}\sim\gamma_{\bar{A}\bar{B}}\sim -1/\log \mu$ for $\mu\rightarrow 0^+$, implying that the integral in eq. \eqref{eq_int_condition} diverges logarithmically for $m_{IR}\rightarrow 0$.
Eq. \eqref{eq_g_condition} is a generic condition on RG flows, which is more general than asymptotic freedom. Provided this is satisfied, the crucial property which ensures \eqref{eq_no_SSB_cond} is the positivity of the anomalous dimension matrix \eqref{eq_anomalous}. For NLSMs with compact target space, this ultimately follows from the unitarity of the model, which is an important assumption of Coleman's theorem as we commented earlier. In more general non-unitary models one has to check if this condition is satisfied case by case.

An analysis similar to the one above applies to higher-point functions of the order parameter.  A famous result for two-dimensional NLSM is that a correlation function is infrared finite at a finite order in perturbation theory only if it is $G$-invariant \cite{Jevicki:1977zn,Elitzur:1978ww,David:1980gi,David:1980rr}. As in the case of one-point functions,  this implies that only $G$-invariant correlators survive in the limit $m_{IR}\rightarrow 0$.\footnote{Let us prove this explicitly for two-point functions. Consider two operators defined as in eq.~\eqref{eq_order_parameter} transforming in two different representations $\tilde{R}$ and $R$. It is simple to verify that their two-point function satisfies:
\begin{equation}\label{eq_2pt_mod_footnote}
\langle\hat{\tilde{\mO}}_A(y)\hat{\mO}_B(0)\rangle=
\langle \hat{\mO}_{AB}\rangle\times G_{AB}(|y|\mu)\,,
\end{equation}
where $G_{AB}(|y|\mu)$ is independent of the IR regulator $m_{IR}$ while the prefactor is the expectation value of the following operator, in obvious notation,
\begin{equation}
 \hat{\mO}_{AB}=\left\{\exp\left[i \pi^a\left(Q^{(\tilde{R})}_a\otimes\mathds{1}+\mathds{1}\otimes Q^{(R)}_a\right)\right]\right\}_{AC,BD}\tilde{v}_C v_D\,.
\end{equation}
The expectation value $\langle \hat{\mO}_{AB}\rangle$ vanishes unless the trivial representation is contained in the tensor product of the representations $\tilde{R}$ and $R$. } It is also possible to check that reflection positivity and clustering are satisfied. Consider for concreteness the connected two-point function $\langle \mO_A^\dagger(x)\mO_B(0)\rangle_c$.  Computing the correlator expanding the exponential in eq.~\eqref{eq_order_parameter} to linear order and using Wick theorem, one naively finds that the tree-level correlation function grows with the distance due to the propagator of the Goldstone fields as in eq.~\eqref{eq_log}.  As well known, this approximation is inaccurate at large distances.  Solving the Callan-Symanzik equation\footnote{In the Abelian case one can equivalently resum all nonlinearities using the Baker-Campbell-Hausdorff formula.} in the perturbative regime, one finds again that the positivity of the anomalous dimension matrix \eqref{eq_gamma_pos} ensures that the correlation function remains positive and decays to zero at large separations provided the condition \eqref{eq_g_condition} is satisfied \cite{MCKANE1980169}.

\subsection{Examples}\label{subsec_CMW_ex}

The simplest and most studied example of NLSM is the $O(N)$ sigma model
\begin{equation}\label{eq_O(N)_NLSM}
S=\frac{1}{g^2}\int d^2x \left(\frac12\pd_\mu\vec{n}\cdot\pd^\mu\vec{n}
-\vec{h}\cdot\vec{n}\right)
\,,\qquad
\vec{n}\cdot\vec{n}=1\,,
\end{equation}
where $\vec{h}=(h,\vec{0})$ is an explicit breaking term analogous to eq. \eqref{eq_L_IR}.
At a classical level $\vec{n}$ is an $O(N)$ vector acting as order parameter for the breaking of the $O(N)$ symmetry to $O(N-1)$
\begin{equation}\label{eq_n_fluct}
\vec{n}= \left(\sqrt{1-\vec{\pi}^2},\vec{\pi}\right)= \left(1,\vec{0}\right) +O(\vec{\pi})\,,
\end{equation}
where the Goldstones $\vec{\pi}$ transform as vectors of the unbroken $O(N-1)$ group.  To one-loop order, the model \eqref{eq_O(N)_NLSM} is made finite introducing wave-function renormalizations for the coupling $g\rightarrow Z_g g$ and  the field $\vec{n}\rightarrow Z_{\vec{n}}\vec{n}$ \cite{Brezin:1976qa}.  We find
\begin{align}\label{eq_Z_O(N)}
Z_{g}=1-\frac{N-2}{4\pi} g^2\log(\Lambda/\mu)\,,\qquad
Z_{\vec{n}}=1-\frac{N-1}{4\pi}g^2\log(\Lambda/\mu)\,,
\end{align}
where $\Lambda$ is the cutoff. Requiring that bare quantities stay invariant, we obtain the well known beta function of the coupling
\begin{equation}\label{eq_beta_O(N)}
\beta_g=-Z^{-1}_g\frac{\pd Z_g}{\pd\log\mu}g=-\frac{N-2}{4\pi} g^3\,,
\end{equation}
as well as the anomalous dimension of the order parameter $\vec{n}$
\begin{equation}\label{eq_anomalous_O(N)}
\gamma_{\vec{n}}=\frac{N-1}{4\pi}g^2\,.
\end{equation}

The solution of the beta function eq. \eqref{eq_beta_O(N)} reads
\begin{equation}\label{eq_RG_O(N)}
g^2(\mu)=\frac{g^2(\mu_0)}{1-\frac{N-2}{2\pi}g^2(\mu_0)\log(\mu_0/\mu)}\,.
\end{equation}
For $N>2$ the coupling diverges at a strong coupling scale
\begin{equation}\label{eq_Lambda_O(N)}
\Lambda_{IR}=\exp\left[-\frac{2\pi}{g^2(\mu_0)(N-2)}\right]\,.
\end{equation} 
No perturbative analysis is therefore possible for $h\rightarrow 0$.
The explicit solution of this model in the large $N$ limit \cite{Moshe:2003xn} or via integrability \cite{Zamolodchikov:1977nu,Zamolodchikov:1978xm} confirms that the theory is gapped and no SSB occurs.

For $N=2$ the theory is instead described by the free compact boson CFT.\footnote{As well known, for sufficiently large $g$ the free boson CFT assumes that relevant interaction terms involving vortex operators are tuned to zero.} The order parameter one-point function as a function of the external magnetic field reads 
\begin{equation}
\langle n_1\rangle\simeq \left(\frac{h}{\mu_0^2}\right)^{\gamma_{\vec{n}}/2}\xrightarrow{h\rightarrow 0}0\,.
\end{equation}
The order parameter two-point function for $h=0$ reads
\begin{equation}
\langle n_A(x) n_B(0)\rangle\propto\frac{\delta_{AB}}{x^{2\gamma_{\vec{n}}}}\,.
\end{equation}

Finally it is also interesting to consider the analytic continuation of the model \eqref{eq_O(N)_NLSM} to $N<2$.  This is expected to describe the low temperature phase of certain loop models and was studied, e.g., in
\cite{Jacobsen:2002wu,Nahum_loop}. A priori, Coleman's theorem does not apply since the NLSM is non-unitary for non-integer $N$ \cite{Binder:2019zqc}. In this case the coupling \eqref{eq_RG_O(N)} approaches zero in the infrared and we can analyze the theory perturbatively.   In agreement with the discussion around eq.~\eqref{eq_g_condition}, we find that no symmetry breaking occurs for $N>1$, since the anomalous dimension \eqref{eq_anomalous_O(N)} is positive.
Indeed from the Callan-Symanzik equation we find 
\begin{equation}\label{eq_n_1pt}
\langle n_1\rangle=\left[1+\frac{2-N}{2\pi}g^2(\mu_0)\log(\mu_0/\sqrt{h})\right]^{-q/2}\,,\qquad
q=\frac{N-1}{2-N}
\end{equation}
which vanishes for $1<N<2$ when $h\rightarrow 0$. Similarly,  correlation functions of the order parameter decay logarithmically with the distance 
\begin{equation}\label{eq_n_2pt}
\langle n_A(x) n_B(0)\rangle\stackrel{|x|\rightarrow\infty}{\propto}
\frac{\delta_{AB}}{\left[\log (|x|\mu_0)\right]^{q}}\,.
\end{equation}

A fully analogous discussion applies to other NLSMs for which the metric is fixed up to an overall factor. For instance, one may consider a coset of the form $(G\times G)/G$, where $G$ is an arbitrary compact group. As well known from the chiral Lagrangian in QCD, the action is constructed out of an element $\Omega[\pi]\in G$  as
\begin{equation}\label{eq_S_GG_NLSM}
S=-\frac{1}{2 \lambda}\int d^2x\text{Tr}\left[\left(\Omega^{-1}\pd_\mu\Omega \right)\left(\Omega^{-1}\pd^\mu\Omega\right)\right]=\frac{1}{2 \lambda}\int d^2x D_\mu\pi^a D^\mu\pi^b\delta_{ab}\,,
\end{equation}
where we assumed the normalization $\text{Tr}\left[Q_a Q_b\right]=\delta_{ab}$ for the generators and $D_\mu\pi^a$ is given in eq.~\eqref{eq_D_A_CCWZ}. A simple calculation shows that the curvature is 
\begin{equation}
\mathcal{R}_{ab}=\frac14 f_{acd}f_{bcd}=\frac{1}{4}\delta_{ab} C_{Adj}\,,
\end{equation}
where $C_{Adj}>0$ is the Casimir of the Adjoint representation. The beta-function~\eqref{eq_beta_NLSM} gives
\begin{equation}\label{eq_beta_GG}
\frac{\pd \lambda}{\pd \log\mu}=-\frac{C_{Adj}}{8\pi}\lambda^2\,,
\end{equation}
and the model is always asymptotically free.

A less trivial example is provided by a fully broken $SO(3)$ group.  The $SO(3)/\emptyset$ Goldstones can be thought as a higher dimensional generalization of the Euler angles describing the rotation of a rigid body, with the covariant derivatives defined in eq.~\eqref{eq_D_A_CCWZ} identified with the components of the angular velocity. As well known from classical mechanics, it is always possible to diagonalize the inertia tensor (via a right action on the coset \eqref{eq_coset}). Thus we can take the NLSM metric to be diagonal at the origin:
\begin{equation}\label{eq_SO(3)_g}
g_{ab}(0)=\text{diag}\left(\frac{1}{k_{1}},\frac{1}{k_{2}},\frac{1}{k_{3}}\right)\,.
\end{equation}
Using that the structure constants are $f_{abc} = \epsilon_{abc}$, in the parametrization \eqref{eq_coset} the Lagrangian reads
\begin{align}\nonumber
    \mathcal{L} = &\frac{1}{2k_{a}}\partial_{\mu}\pi^{a}\partial^{\mu}\pi^{a}
    - \frac{1}{2k_{a}}\epsilon_{cab}\pi^{c}\partial_{\mu}\pi^{a}\partial^{\mu}\pi^{b}
    + \frac{1}{8k_{e}}\epsilon_{ace}\epsilon_{bde}\pi^{c}\pi^{d}\partial_{\mu}\pi^{a}\partial_{\mu}\pi^{b}\\
    & - \frac{2}{3k_{a}}\pi^b\pi^b\partial_{\mu}\pi^{a}\partial^{\mu}\pi^{a}
    + \frac{2}{3k_{a}}\pi^{a}\pi^{b}\partial_{\mu}\pi^{a}\partial^{\mu}\pi^{b}
    + O(\pi^{3}\partial\pi\partial\pi)
\end{align}

The running of the three coupling constants $k_a$ is governed by the curvature. We find
\begin{align}\label{eq_SO(3)_R}
    \mathcal{R}^{ab} = 
     \begin{pmatrix}
        \frac{k_{2}^{2}k_{3}^{2}-k_{1}^{2}(k_{2}-k_{3})^{2}}{2k_{2}k_{3}} & 0 & 0\\
        0 & \frac{k_{1}^{2}k_{3}^{2}-k_{2}^{2}(k_{1}-k_{3})^{2}}{2k_{1}k_{3}} & 0 \\
        0 & 0&  \frac{k_{1}^{2}k_{2}^{2}-k_{3}^{2}(k_{1}-k_{2})^{2}}{2k_{1}k_{2}}
    \end{pmatrix}\,.
\end{align}
Notice that this is also diagonal. The beta functions then read
\begin{equation}
\frac{\partial k_{a}}{\partial \log\mu} = -\frac{1}{2\pi}\mathcal{R}^{aa}\qquad\text{(no sum)}\,.
\end{equation}
In agreement with the general analysis, the curvature $\mathcal{R}^{ab}$ is never negative definite. To see this, assume first that the top two diagonal entries are negative, which gives the inequalities 
\begin{align}
    k_{1}^{2}(k_{2}-k_{3})^{2} > k_{2}^{2}k_{3}^{2}\,,
    \hspace{35pt}
    k_{2}^{2}(k_{1}-k_{3})^{2} > k_{1}^{2}k_{3}^{2}\,.
\end{align}
Adding these two equations, we obtain the following inequality
\begin{align}
    k_{1}k_{2} > k_{3}(k_{1}+k_{2}) \geq k_{3}|k_{1}-k_{2}|\,,
\end{align}
which implies the third diagonal entry must be positive.  Thus the couplings run to strong values and the theory is not perturbative in the IR.  

\begin{figure}[t!]
\centering
\includegraphics[scale=0.7]{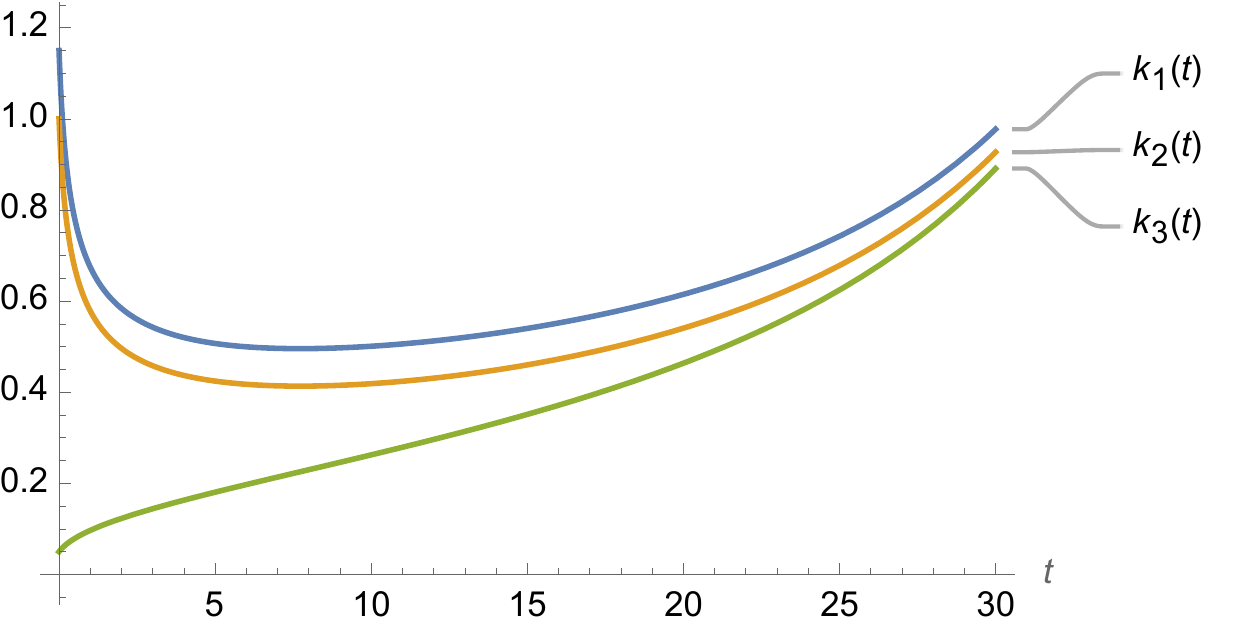}
\caption{Plot of the running of coupling constants in the fully broken $SO(3)$ example for initial conditions such that two entries of the matrix~\eqref{eq_SO(3)_R} are negative. The horizontal axis is the RG time $t = -\log(\mu/\mu_{0})$.}
\label{fig:so}
\end{figure}

It is also interesting to consider the RG evolution of the differences between coupling constants. A simple calculation shows
\begin{align}
    &\frac{\partial(k_{1}-k_{2})}{\partial\log\mu} = \frac{1}{2\pi} (k_{1}-k_{2})\frac{(k_{1}k_{2}-k_{1}k_{3}-k_{2}k_{3})^{2}}{2k_{1}k_{2}k_{3}}\,,\\
    &\frac{\partial(k_{1}-k_{3})}{\partial\log\mu} = \frac{1}{2\pi}(k_{1}-k_{3})\frac{(k_{1}k_{3}-k_{1}k_{2}-k_{2}k_{3})^{2}}{2k_{1}k_{2}k_{3}}\,.
\end{align}
This shows the running of the differences $k_{a}-k_{b}$ is always proportional to itself multiplied by a positive coefficient. As a consequence, for any two couplings, the sign of $k_{a}-k_{b}$ will stay fixed, and they will approach each other under RG. As they become close, both $g_{ab}$ and $\mathcal{R}^{ab}$ become proportional to $\delta_{ab}$ with a positive proportionality coefficient, as required by eq.~\eqref{eq_RG_Einstein}.  In fig.~\ref{fig:so} we show a plot of the generic running of the coupling constants.

\section{Spontaneous symmmetry breaking on defects and Ward identities}\label{sec_SSB_WI}

\subsection{Ligthning review of DQFTs}\label{subsec_DQFT_review}

In this section we review some aspects of defect QFTs (DQFTs) that will be relevant for our discussion. We refer the reader to \cite{McAvity:1993ue,Jensen:2015swa,Billo:2016cpy,Herzog:2017xha,Herzog:2020bqw,Cuomo:2021cnb} for more details.

Let us consider a $d>2$-dimensional CFT  invariant under a symmetry group $\tilde{G}$. Bulk operators are labeled as usual according to their quantum numbers under the conformal group $SO(d+1,1)$ and the continuous internal group $\tilde{G}$.  These include in particular $n_{\tilde{G}}$ conserved currents with scaling dimension $\Delta_J=d-1$. 

We now introduce a planar $p$-dimensional defect or boundary $\mathcal{D}$ placed at $z^m_\bot=0$, in terms of the coordinates $x^\mu=(y^m,z_\bot^k)$, where $m=1,2,\ldots,p$ and $k=p+1,\ldots,d$.  The defect in general breaks \emph{explicitly} the bulk conformal symmetry as well as the internal symmetry group to the $p$-dimensional Euclidean group and an internal subgroup $G\subset \tilde{G}$; often, the transverse rotations $SO(d-p)$ are preserved as well.  Local defect operators are labeled by their quantum numbers under the symmetry group preserved by the defect.

Bulk conservation laws are modified by defect operators.  In particular, the conservation of the Noether currents $J_{\rm a}^\mu$ is generally modified by a scalar defect operator $\hat{t}_a(y)$ as \cite{Metlitski:2020cqy,Cuomo:2021cnb,Padayasi:2021sik}
\begin{equation}\label{eq_current_defect}
\pd_\mu J^\mu_{\rm a}(x)=\delta^{d-p}(z_\bot)\hat{t}_{\rm a}(y)\,.
\end{equation}
It is important to distinguish two cases:
\begin{enumerate}
\item The operator $\hat{t}_{\rm a}(y)$ is either zero or a total derivative:
\begin{equation}\label{eq_defect_currents}
\hat{t}_{\rm a}(y)=-\pd_m\hat{J}^m_{\rm a}(y)\,,
\end{equation}
where the operators $\hat{J}^m_{\rm a}$, when they exist, are referred to as the defect currents.
In this case, the DQFT admits a topological operator given by
\begin{equation}\label{eq_charge}
Q_{\rm a}=\int_{\Sigma}\star J_{\rm a}+\int_{\Sigma\cap\mathcal{D}}\star\hat{J}_{\rm a}\,,
\end{equation}
where $\Sigma$ is an arbitrary $d-1$-dimensional surface and $\Sigma\cap\mathcal{D}$ denotes the intersection of this surface with the defect. The topological operators constructed in this way generate the group $G\subset\tilde{G}$ left unbroken by the defect.   Obviously, when the defect current does not exist, the second contribution on the right hand side of eq.~\eqref{eq_charge} vanishes. This however does not mean that the generator~\eqref{eq_charge} acts trivially on local defect operators.
\item The operator $\hat{t}_{\rm a}(y)$ is not a total derivative. In this case one cannot construct a topological operator as in eq.~\eqref{eq_charge} when the surface $\Sigma$ intersects the defect. The $n_{\tilde{G}}-n_{G}$ operators $\hat{t}_{\rm a}(y)$ which are not total derivatives are sometimes referred to as \emph{tilt} operators. They always exist when a continuous symmetry is \emph{explicitly} broken by the defect.
\end{enumerate}

Notice that when eq.~\eqref{eq_defect_currents} holds, the Ward identity~\eqref{eq_current_defect} is left unchanged by the transformation
\begin{equation}\label{eq_ambiguity}
\hat{J}^m_{\rm a}(y)\rightarrow(1+a)\hat{J}^m_{\rm a}(y)\quad\text{with}\quad
J^\mu_{\rm a}(x)\rightarrow J^\mu_{\rm a}(x)-a \delta^{d-p}(z_\bot)\delta^\mu_m\hat{J}^m_{\rm a}(y)\,.
\end{equation}
This makes generally ambiguous the distinction between bulk and defect current.\footnote{We thank P. Kravchuk for useful discussions about this.} As we will see, there are nonetheless situations where this ambiguity is resolved by additional constraints.

We now discuss in more detail the case in which the defect is conformal, i.e. it preserves an $SO(p+1,1)$ subgroup of the bulk conformal group. This setup defines a defect CFT (DCFT). In this case defect operators form nontrivial representations of the $p$-dimensional conformal group.  
Additionally, in DCFTs bulk operators can be expanded as sum of defect operators via the bulk-to-defect OPE, which schematically takes the form (in obvious notation)
\begin{equation}
\mO(x)\sim\sum_{\hat{\mO}}c_{\mO\hat{\mO}}|z_{\bot}|^{\hat{\Delta}_{\hat{\mO}}-\Delta_{\mO}}\hat{\mO}(y^m)\,,
\end{equation}
where we neglected spin indices and the coefficients $c_{\mO\hat{\mO}}$ are theory-dependent (up to the constraints of conformal invariance).

An important observation is the following: in DCFTs,  generators preserved by the defect never admit defect currents. This is because the identities \eqref{eq_current_defect} and \eqref{eq_defect_currents} imply that the defect currents, when they exist, are primary operators of spin $J=1$ and scaling dimension $\hat{\Delta}_{\hat{J}}=p-1$.\footnote{Notice that the Ward identity~\eqref{eq_current_defect} implies that $\hat{t}_{\rm a}$ transforms as a primary of the $p$-dimensional conformal group. This thus rules out the possibility that $\hat{J}^m_{\rm a}$ is a descendant of another operator. One can also prove that a conserved bulk current admits a nontrivial two-point function with a defect scalar primary operator only if the scalar has dimension $p$ \cite{Herzog:2020bqw}.} However, the representation theory of the $p$-dimensional conformal group implies that primaries with $\Delta-J=p-2$ and $J\geq 1$ are conserved, and therefore a defect current cannot contribute nontrivially to eq. \eqref{eq_current_defect} since $\hat{t}_{\rm a}(y)=\pd_m\hat{J}^m_{\rm a}(y)=0$.  

Conserved defect currents, if they exist, unavoidably generate a symmetry group $G'$ which is different than the one generated by the bulk Noether currents. Since $G'$ is not a symmetry of the bulk CFTs, only defect operators can be charged under the latter.
In particular, in DCFTs, operators charged under $G'$ cannot appear in the bulk-to-defect OPE of bulk operators.  
We will see that at the end of certain defect RG flows, defect currents may emerge along with a decoupled sector of charged operators.

The Ward identity \eqref{eq_current_defect} also implies that, in DCFT, tilt operators associated with generators not preserved by the defect have dimension $\hat{\Delta}_{\hat{t}}=p$, and thus provide marginal defect deformations.\footnote{To the best of our knowledge, the existence of an operator with scaling dimension exactly $p$ was first noticed in \cite{Bray_1977,Ohno} in the context of the extraordinary boundary universality class in the $O(N)$ model.} Indeed, there generally are several equivalent ways in which the subgroup $G$ preserved by the defect can be chosen. Correspondingly the DCFT admits a defect conformal manifold, isomorphic to the coset $\tilde{G}/G$, generated by the tilt operators \cite{Drukker:2022pxk,Herzog:2023dop}.\footnote{Another argument for the existence of $n_{\tilde{G}}-n_G$ defect marginal operators is as follows. We can use a Weyl rescaling to map the DCFT to AdS$_{p+1}\times S^{d-p-1}$, with the defect sitting at the boundary of AdS$_{p+1}$. In these coordinates,  bulk operators have constant expectation value and the internal symmetry is spontaneously broken in the bulk of AdS$_{p+1}$. Thus there exist $n_{\tilde{G}}-n_G$ massless bulk fields, the Goldstones.  As well known from the standard AdS/CFT dictionary, the lowest component in the near boundary expansion of a massless field in AdS corresponds to a marginal operator at the boundary. Upon performing a KK reduction on the sphere, the lowest components of the AdS$_{p+1}$ Goldstones are precisely the \emph{tilt} operators.
\label{footnote_tilt}}

\subsection{Spontaneous symmetry breaking on defects}\label{subsec_SSB_defects}

Consider a DQFT invariant under a continuous internal symmetry $G$, i.e.  $G$ is a symmetry group which is preserved by the defect.  In this section we only consider defects of dimension $p\geq 2$; we will comment on line defects at the end of sec.~\ref{sec_free_ex}. Suppose that the ground-state breaks spontaneously this symmetry group, i.e. there exists a local defect operator which transforms in a (not necessarily irreducible) representation $R$ of $G$ whose expectation value is not invariant under the action of the internal symmetry group
\begin{equation}\label{eq_SSB_defect}
\langle\hat{\mO}_A\rangle= v_A\quad\text{such that}\quad \langle [Q_a,\hat{\mO}_A]\rangle = -\delta_a v_A\neq 0\,,
\end{equation}
where the hat distinguishes defect operators from bulk operators and the rest of the notation is analogous to that of sec.~\ref{subsec_CMW_thm}.  Eq. \eqref{eq_SSB_defect} in general implies, via the bulk-to-defect OPE, that bulk operators also acquire an expectation value breaking the symmetry. We assume however that the symmetry is restored at infinite distance from the defect, i.e.
\begin{equation}\label{eq_SSB_defect_condition2}
\lim_{z_\bot\rightarrow\infty}\langle\mO_A(y,z_\bot)\rangle=0
\end{equation}
for all bulk operators. 

We want to discuss the consequences of eq.~\eqref{eq_SSB_defect} for the DQFT. The generalization of the Ward identity \eqref{eq_WI_bulk} to defects, in general, involves both the bulk and the defect currents
\begin{equation}\label{eq_WI_defect}
\langle\left(\pd_\mu J^\mu_a(y,z_\bot)+\delta^{d-p}(z_\bot)\pd_m\hat{J}^m_a(y)\right)
\hat{\mO}_A(0)\rangle= -\delta_a v_A\delta^{d-p}(z_\bot)\delta^p(y)\,.
\end{equation} 
Eq.~\eqref{eq_WI_defect} implies
\begin{equation}\label{eq_defect_pole}
\int d^pye^{iky}\langle Y^m_a(y)
\hat{\mO}_A(0)\rangle=-i\delta_a v_A\frac{k^m}{k^2}+i\varepsilon^{mn}k_n f(k^2)\,,
\end{equation}
where the last term exists only for $p= 2$ and we defined the following nonlocal operator
\begin{equation}
Y^m_a(y)=\hat{J}^m_a(y)+\int d^{d-p}z_{\bot}J^m_a(y,z_\bot)\,.
\end{equation}
Eq.~\eqref{eq_defect_pole} is analogous to eq.~\eqref{eq_Hmu} in QFT. The main difference between a DQFT and a QFT is that the operator $Y^m_a$ whose two-point function with the order parameter admits a pole for $k^2\rightarrow 0$ is nonlocal. As we commented below eq.~\eqref{eq_H_rho2}, we therefore cannot conclude, in general, that eq.~\eqref{eq_defect_pole} implies the existence of a massless pole at zero-momentum in the two-point function of the order parameter.

It is useful to discuss this point in more detail.  Let us consider the theory in Lorentzian signature, with the time coordinate parallel to the defect. We assume the existence of a Hilbert space on constant time slices. It follows from the residual translational symmetry that the states of such Hilbert spaces can be labeled by their momentum in the directions parallel to the defect, along with other quantum numbers that we do not need to specify.

Under this assumption, we can derive the spectral decomposition of an arbitrary two-point function by inserting a complete set of states between the operators, similarly to eq.~\eqref{eq_H_rho}.  For the correlator of the bulk current with the defect order parameter we find:  
\begin{equation}
\langle J^m_a(y,z_\bot) \hat{\mO}_A(0)\rangle=
\int_{0}^\infty dm^2\int\frac{d^pk}{(2\pi)^p} e^{-i k y}\frac{k^m\rho_{bulk}(m^2,z_\bot)}{k^2+m^2}\,,
\end{equation}
where the spectral density is defined as
\begin{equation}\label{eq_rho_bulk_defect}
k^m\rho_{bulk}(k^2,z_\bot)=
(2\pi)\sum_n\langle 0|J_a^m(0,z_\bot)|n\rangle\langle n|\hat{\mO}_A(0)|0\rangle\delta^p(k-q_n)\,.
\end{equation}
Notice that, due to the lack of translational invariance in the directions perpendicular to the defect, the spectral density depends explicitly on the transverse coordinates $z_\bot$.
The spectral decomposition of the defect current correlator (when it exists) is analogous to the standard one in eq.~\eqref{eq_H_rho}:
\begin{equation}\label{eq_rho_j_defect}
\langle \hat{J}^m_a(y) \hat{\mO}_A(0)\rangle=
\int_{0}^\infty dm^2\int\frac{d^pk}{(2\pi)^p} e^{-i k y}\frac{k^m\hat{\rho}_{P}(m^2)}{k^2+m^2}\,.
\end{equation}
For specific values of $d$ and $p$ there exist additional transverse contributions in eq.~\eqref{eq_rho_bulk_defect} and eq.~\eqref{eq_rho_j_defect}; these contributions do not play a role in our analysis. Eq.~\eqref{eq_defect_pole} thus implies
\begin{equation}\label{eq_rho+rho_defect}
\hat{\rho}_P(m^2)+\int d^{d-p}z_{\bot}\rho_{bulk}(m^2,z_\bot)=-i\delta_a v_A\delta(m^2)\,.
\end{equation}

This equation may be satisfied in several different ways. If the delta function arises from the defect spectral density, we can run the usual argument and conclude that there exist massless states whose matrix elements with the order parameter are non-trivial.   Indeed,  as previously remarked,  the eq. $\hat{\rho}_P(m^2)= -i\delta_a v_A\delta(m^2)+\ldots$ cannot be satisfied without the existence of massless particles admitting a nontrivial matrix element with the order parameter.
Therefore when the defect correlator admits a pole for $k^2\rightarrow 0 $, also the two-point function of the order parameter displays a pole at zero-momentum; this in turn leads to an inconsistency for surface defects $p=2$ as in the usual Coleman's theorem.

Alternatively, it is possible that $\hat{\rho}_P(m^2)$ does not contain any delta function at $m^2=0$ and
\begin{equation}\label{eq_rho_int_defect}
\int d^{d-p}z_{\bot}\rho_{bulk}(m^2,z_\bot)=-i\delta_a v_A\delta(m^2)+\ldots\,,
\end{equation}
where the dots schematically denote additional terms that are needed to satisfy eq.~\eqref{eq_rho+rho_defect}. Since what appears in eq.~\eqref{eq_rho_int_defect} is the \emph{integrated} spectral density, the constraint~\eqref{eq_rho_int_defect} may safely be saturated by the continuum part of the spectrum without implying the existence of singular matrix elements for the current $J^\mu_a$ or the order parameter $\hat{\mO}_A$.\footnote{A simple example of how eq.~\eqref{eq_rho_int_defect} can be satisfied by a continuum spectrum is provided by a local QFT which breaks spontaneously $G\rightarrow H$ in dimension $d>2$.  We may formally regard the local QFT as a DQFT in the presence of a trivial defect of dimension $p<d$.  In this case, the defect order parameter is the usual order parameter $\mO_A(y,z_\bot)$ at $z_\bot=0$.  From the viewpoint of the defect, the bulk Goldstone bosons contribute as a continuum of states with mass given by their transverse momentum $m^2=k_\bot^2$.  Correspondingly, one can check that the spectral density at fixed $z_\bot$ defined in eq.~\eqref{eq_rho_bulk_defect} reads 
\begin{align}
\rho_{bulk}(m^2,z_\bot)&=-i\delta_av_A\int \frac{d^{d-p}k_{\bot}}{(2\pi)^{d-p}}e^{i k_{\bot}\cdot z_\bot}\delta(m^2-k_\bot^2)\, \nonumber\\
&=\frac{-i\delta_{a}v_{A}}{2(2\pi)^{\frac{1}{2}(d-p)}}\left(\frac{m}{|z_{\perp}|}\right)^{\frac{1}{2}(d-p-2)}J_{\frac{1}{2}(d-p-2)}(m|z_{\perp}|)\,,
\end{align}
which indeed satisfies eq.~\eqref{eq_rho_int_defect}.  Notice that this setup violates the condition~\eqref{eq_SSB_defect_condition2} and thus does not provide an example of defect symmetry breaking when specialized to $p=2$. We will construct an example of symmetry breaking on a two-dimensional defect which satisfies both conditions~\eqref{eq_SSB_defect} and \eqref{eq_SSB_defect_condition2} in sec.~\ref{sec_free_ex}.}

It may thus seem that little mileage can be gained from Ward identities when one considers SSB on surface defects. In particular, there seem to be no obstruction for the SSB of continuous internal symmetries on surface defects. We will indeed construct such an example in sec.~\ref{sec_free_ex}.  An important feature of this example is that the DQFT never becomes conformal nor scale invariant even at arbitrary large distances.

We now want to discuss theories in which the defect RG terminates in a well behaved fixed point. What we mean by this is that correlation functions at large distances are well approximated by those of a scale invariant theory up to the (calculable) effects of irrelevant deformations.  We also assume that the scale invariant theory describing the IR limit admits a diagonalizable dilation operator with a discrete spectrum, and that bulk operators retain their UV scaling dimensions at such fixed-point.

We expect the above conditions to be met in most RG flows of interest. In fact,  these conditions hold in local QFTs with SSB, in which case the IR theory generically consists of free massless Goldstone fields. One generically expects the IR DQFT to further enjoy full $p$-dimensional conformal invariance. It is known however that, when dealing with SSB, the issue of scale vs conformal invariance becomes a subtle one (and in some sense a semantic one).\footnote{Consider indeed a (local) theory for the Goldstone fields parametrizing the coset $G/H$ in dimensions $d>2$. On the one hand, all correlation functions at large distances are well approximated by those of a free theory. On the other hand, it is known that the improvement term which is required to make the stress tensor traceless is incompatible with the nonlinearly realized symmetry.   See \cite{Luty:2012ww,Nakayama:2013is,Dymarsky:2013pqa} for a more detailed analysis of this example. We thank Z.Komargodski for useful discussions on this.} We will only need the assumptions stated above.

Under our assumptions,  we can compute all correlation functions in terms of the operators that diagonalize the infrared dilation operator. Bulk operators retain the same scaling dimension along the defect RG flow by assumption.  Correlation functions of defect operators are reproduced by a sum of operators with definite IR scaling dimension. Consider for instance the order parameter. The VEV~\eqref{eq_SSB_defect} implies that the defect order parameter flows to a sum of defect operators that includes the identity:
\begin{equation}\label{eq_O_flow}
\hat{\mO}_A\xrightarrow{IR}v_A
\hat{\mathds{1}}+\sum_{\Delta\geq 0} c_{AB}^{(\Delta)} \hat{\mO}_{B}^{(\Delta)}\,,
\end{equation}
where the coefficients $c_{AB}^{(\Delta)}$ may be constrained by the internal symmetry.   Notice that all scaling dimensions must be positive due to the cluster decomposition principle.  Eq.~\eqref{eq_O_flow} implies that the long distance limit for the connected correlators of the order parameter is reproduced by the contribution of the first non-trivial operator in eq.~\eqref{eq_O_flow}, whose scaling dimension we denote $\Delta_1$. For instance, the  two-point function of the order parameter at large separations is reproduced by 
\begin{equation}
\langle \hat{\mO}_A(y)\hat{\mO}_B(0)\rangle_c\stackrel{y\rightarrow\infty}{=}
 c_{AC}^{(\Delta_1)}c_{BD}^{(\Delta_1)} \langle\hat{\mO}_{B}^{(\Delta_1)}(y)\hat{\mO}_{D}^{(\Delta_1)}(0)\rangle_{IR}+\ldots
\end{equation}
where $\langle\hat{\mO}_{B}^{(\Delta_1)}(y)\hat{\mO}_{D}^{(\Delta_1)}(0)\rangle_{IR}\sim 1/|y|^{2\Delta_1}$ denotes the connected correlator within the IR scale invariant theory and the dots denote terms which decay faster than $1/|y|^{2\Delta_1}$ for $y\rightarrow\infty$. As an example, in the case of SSB in local $d>2$-dimensional QFTs, from eq.~\eqref{eq_order_parameter} we see that the first nontrivial contribution in eq.~\eqref{eq_O_flow} arises from a free Goldstone field $\pi^a$,  which has dimension $\Delta_1=(d-2)/2$.

Let us work for simplicity with the order parameter normalized so that its expectation value $v_A$ is dimensionless,  as it can always be achieved rescaling $\hat{\mO}_A\rightarrow\hat{\mO}_A/\sqrt{\sum_B v_B^* v_B}$.  Our main observation is that eq.~\eqref{eq_O_flow} implies that the two-point function of the bulk current and the order parameter cannot lead to a delta function upon conservation. Indeed, by dimensional analysis, for $\langle\pd_\mu J^\mu_a(x)\hat{\mO}_A(0)\rangle\supset- \delta_a v_A\delta^d(x)$ to be satisfied the order parameter $\hat{\mO}_A$ would need to include a nontopological dimensionless operator in the IR decomposition~\eqref{eq_O_flow}.\footnote{Notice that in a local QFT, SSB leads to a free Goldstone sector in the IR, with the internal currents flowing to the free field shift currents. These thus have IR dimension $d/2$, while the order parameter includes a dimension $(d-2)/2$ free Goldstone field in eq.~\eqref{eq_O_flow}. This allows satisfying the Ward identity \eqref{eq_WI_bulk}.}  However, in a scale invariant theory with discrete spectrum of dimensions, a dimension $0$ operator $\hat{\mO}^{(0)}$ must have a constant two-point function $\langle\hat{\mO}^{(0)}(y)\hat{\mO}^{(0)}(0)\rangle=\text{const.}$. 
This implies that the derivatives $\pd^k \hat{\mO}^{(0)}$ act trivially on the vacuum. Thus $\hat{\mO}^{(0)}$ must be topological, in the sense that its correlation functions cannot depend on its insertion point. Therefore a dimension zero operator $\hat{\mO}^{(0)}$ cannot contribute nontrivially in the Ward identity.

For the Ward identity~\eqref{eq_WI_defect} to be satisfied we thus crucially need the contribution of the defect currents.  Notice that these might be only emergent in the IR.  The currents' correlators with the order parameter therefore read
\begin{align} \label{eq_J_O_g}
\langle  J^\mu_a(x)\hat{\mO}_A(0)\rangle &=\frac{\pd^\mu}{\pd_\nu\pd^\nu}\delta^{d-p}(z_\bot)\frac{f_{aA}(|y|\Lambda)}{|y|^p}+\ldots\,,\\
\langle \hat{J}^m_a(y)\hat{\mO}_A(0)\rangle&=-\delta_a v_A\frac{\pd^m}{\pd^n  \pd_n}\delta^p(y)-\frac{\pd^m}{\pd^n\pd_n}\frac{f_{aA}(|y|\Lambda)}{|y|^p}+\ldots\,, \label{eq_J_O_g_2}
\end{align}
where the dots denote additional explicitly conserved terms, which will not play a role in our discussion, and $1/\pd_\mu\pd^\mu$ and $1/\pd_m\pd^m$ denote, respectively,  the $d$- and $p$-dimensional inverse Laplacians. This structure follows from dimensional analysis and bulk conservation. As explained above, our assumptions rule out a contribution of the form $\propto \frac{\pd^\mu}{\pd_\nu\pd^\nu} \delta^d(x)$ in the bulk-current correlator.  Notice that in order to satisfy the Ward identities for the broken generators, the defect currents must have IR scaling dimension different than $p-1$ (in fact, from the discussion below it follows that they have dimension $p/2$ for $p\neq 2$).  Because of the different transformation properties of the bulk and defect currents under the infrared emergent dilations, redefinitions of the form~\eqref{eq_ambiguity} are not compatible with scale invariance. This implies that the IR defect currents are unambiguous and can be clearly distinguished from the bulk currents within our assumptions.

By the assumption of scale invariance, the function $f(|y|\Lambda)$ vanishes when the cutoff is taken to infinity, $f(|y|\Lambda)\xrightarrow{\Lambda\rightarrow\infty}0$. In general we expect $f(|y|\Lambda)\sim (|y|\Lambda)^{-\delta}$ for some $\delta>0$, whose value is related to the dimension of the leading irrelevant deformation. When the RG flow is logarithmic, i.e. the leading deformation is only \emph{marginally} irrelevant,  the function $f(|y|\Lambda)$ might be suppressed by a logarithm of the UV scale $f(|y|\Lambda)\sim (|y|\Lambda)^{-\delta}[\log(|y|\Lambda)]^{-\tilde{\delta}}$ for some $\delta\geq 0$ and $\tilde{\delta}>0$.\footnote{It is not always true that the IR limit of a correlation function in the presence of a marginally irrelevant coupling approaches a scale invariant limit in the IR - sometimes the RG flow leads to (logarithmically) faster decay at large distances, similarly to the correlator in eq.~\eqref{eq_n_2pt}.
These subtleties however do not affect the correlator of the bulk current and the defect order parameter \eqref{eq_J_O_g}, since scale invariance predicts a trivial result for the non-transverse contribution.}

Let us stress that the Ward identity~\eqref{eq_SSB_defect} constrains the correlation functions at arbitrary distances. In particular, it is not possible to modify the first term in eq.~\eqref{eq_J_O_g_2} consistently with it, and thus the two-point function of the defect current and the order parameter must decay as $\sim y^m/|y|^{p}$ for $y\rightarrow\infty$.  Due to the importance of this fact, it is useful to rephrase the argument in terms of the action of the topological operators.  Consider integrating the normal components of the currents in eq.s~\eqref{eq_J_O_g} and~\eqref{eq_J_O_g_2} over the surface of a ball of very large radius $|y|=|z_\bot|=R\gg \Lambda^{-1}$. By the definition of the topological operator $Q_a$ in eq.~\eqref{eq_charge}, the sum of the integrated bulk and defect correlation functions is equivalent to the action of the charge on the order parameter and therefore it must be non-zero. Explicitly,  in obvious notation,
\begin{equation}\label{eq_top_action}
\int_{|x|=R}d^{d-1}\Sigma_{\mu}  \langle  J^\mu_a(x)\hat{\mO}_A(0)\rangle +
\int_{|y|=R}d^{p-1}\hat{\Sigma}_{m}
\langle \hat{J}^m_a(y)\hat{\mO}_A(0)\rangle=-\delta_av_A\,.
\end{equation} 
Notice that we may also deform the contour of the first integral to an arbitrary surface whose intersection with the defect coincides with the ball $|y|=R$.
By Stokes' theorem, only the terms explicitly written in eq.~\eqref{eq_J_O_g} and~\eqref{eq_J_O_g_2} contribute to the integrals in eq.~\eqref{eq_top_action}. By our assumptions the contribution of the first term in eq.~\eqref{eq_top_action} vanishes for $R\rightarrow\infty$, and therefore the integral of the defect current must be non-zero. This implies that the correlator of the defect current and the order parameter decays as $\sim y^m/|y|^{p}$ for $y\rightarrow\infty$ up to explicitly conserved terms.

From the long distance behaviour of the defect currrent correlator~\eqref{eq_J_O_g_2}, we extract the spectral density $\hat{\rho}_P(m^2)$ defined in eq.~\eqref{eq_rho+rho_defect} for $m^2\ll \Lambda^2$.  The first term on the right-hand side of eq.~\eqref{eq_J_O_g_2} leads to a discrete contribution in the defect correlator spectral density $\hat{\rho}_P(m^2)=-i\delta_a v_A\delta(m^2)+\ldots$.  The term $f(|y|\Lambda)$ additionally introduces a contribution from the mass continuum in the spectral density. This contribution is integrable for $m^2\rightarrow 0$ and can be distinguished from the discrete one.  We thus conclude that there exist $n_G-n_H$ \emph{defect} Goldstone bosons interpolated by the order parameter.  

In conclusion, under the assumption of infrared scale invariance, the IR limit of the DQFT must include a decoupled defect sector with the Goldstones.  In particular eq.~\eqref{eq_J_O_g_2} implies that the order parameter two-point function must have a pole at zero momentum; repeating the argument in sec.~\ref{sec_Coleman_review} we thus conclude that SSB cannot occur on surface defects if the defect RG flow terminates in a scale invariant fixed point in the IR. This is the main result of this work.

In the next two sections we will exemplify the abstract arguments of this section. In particular, in sec.~\ref{sec_free_ex} we will discuss an example of symmetry breaking on a surface defect. In sec.~\ref{sec_log} we will instead discuss in detail the generic endpoint of the RG flow for a $p$-dimensional defect which exhibits SSB. We will then specialize to $p=2$ and discuss how the defect version of Coleman's argument is realized. We will find that the structure of perturbative RG flows for NLSMs coupled to DCFTs is richer than in local QFTs.

\section{Free scalar example}\label{sec_free_ex}

Let us consider the following model consisting of $N$ free scalars in $d>2$ dimensions coupled to a $O(N)/O(N-1)$ NLSM on a two-dimensional surface\footnote{Notice that in eq.~\eqref{eq_free_ex} we rescaled $h\rightarrow g^2 h$ with respect to the notation in eq.~\eqref{eq_O(N)_NLSM}.}
\begin{equation}\label{eq_free_ex}
S=\int d^dx\frac{1}{2}(\pd_\mu\vec{\phi})^2+\int_{z_\bot=0} d^2y
\left[\frac{1}{2g^2}(\pd_m\vec{n})^2-h\,\vec{n}\cdot\vec{\phi}\right]\,,\qquad
\vec{n}^2=1\,.
\end{equation}
The coupling $g$ is marginal while $h$ has dimension $3-d/2$ and it is thus relevant for $d<6$.  For $d=6$ the coupling is marginal, while for $d>6$ it is irrelevant. Below we show that for $d<6$ the model does not flow to a DCFT and the vacuum breaks spontaneously the $O(N)$ symmetry.

To understand the peculiarities of this model it is useful to consider first the theory in which the NLSM is not dynamical, i.e. $n_a=\delta_a^1$ and the surface breaks explicitly the symmetry. In this case the defect acts as a trivial source for the scalar field which is expanded as
\begin{equation}\label{eq_h_1pt}
\phi_A(x)=\frac{2\pi}{(d-2)(d-4)\Omega_{d-1}}\frac{h}{|z_\bot|^{d-4}}\delta_A^1+\delta\phi_A(x)\,,
\end{equation}
where $A=1,2,\ldots,N$, $\Omega_{d-1}=\frac{2\pi^{d/2}}{\Gamma(d/2)}$ is the volume of the $d$-dimensional sphere and $\delta\phi_A(x)$ is a decoupled free field.  The one-point function of the field never approaches a conformal scaling for $d<6$. The parameter $h$ is exactly marginal for $d=6$, in which case the system defines a DCFT.  The field vanishes at infinite distance $z_\bot\rightarrow\infty$ for $d>4$ (in four-dimensions the one-point function grows logarithmically).  

To gain further insights on this \emph{source} defect, in appendix \ref{app_runaway} we compute the partition function for a spherical defect of radius $R$. The universal part $s$ of such partition function provides a $c$-function for the defect RG flow \cite{Jensen:2015swa,Shachar:2022fqk}.  We find that $s\rightarrow-\infty$ as we take the defect radius $R$ to infinity.  This is again in contrast with RG flows that terminate in infrared DCFTs, for which $s$ is finite.  Similar examples of defect RG flows which do not terminate in DCFTs were discussed in \cite{Cuomo:2021kfm,Cuomo:2022xgw} for line defects in free theory.  We call this behaviour a \emph{runaway} defect RG flow.

Physically, such a behaviour is possible because the bulk theory admits a moduli space of vacua. Conformal invariance of the bulk theory allows choosing the dimensionful defect coupling $h$ at will, and thus we can make the expectation value of the field arbitrarily large at long distances. 
We comment further about the relation between source defects and moduli spaces in app.~\ref{app_source_defects}.
In generic interacting theories, we expect to recover conformal invariance at large distances, with one-point functions of bulk operators that do not depend on tunable coefficients.

Restoring the dynamics of he defect field $\vec{n}$ does not change qualitatively the infared behaviour of the DQFT. It is convenient to integrate out explictly the free bulk field on its equations of motion
\begin{equation}
\vec{\phi}(y,z_\bot)=\int d^2y'\frac{\vec{n}(y')}{(d-2)\Omega_{d-1}\left[z_\bot^2+(y-y')^2\right]^{\frac{d-2}{2}}}+\delta\vec{\phi}(y,z_\bot)\,,
\end{equation}
where $\delta\vec{\phi}$ is a free field fluctuation which decouples from the surface. Neglecting $\delta\vec{\phi}$, the theory reduces to the following nonlocal model
\begin{equation}\label{eq_S_nonlocal}
S=\frac{1}{2g^2}\int d^2y\left[(\pd_m \vec{n})^2-
\frac{g^2 h^2}{(d-2)\Omega_{d-1}}
\int d^2y'\frac{\vec{n}(y)\cdot \vec{n}(y')}{|y-y'|^{d-2}}\right]\,.
\end{equation}

The study of the model \eqref{eq_S_nonlocal} is analogous to the analysis of a spin impurity in free theory (Bose-Kondo model) discussed in \cite{Beccaria:2022bcr,Cuomo:2022xgw,Nahum:2022fqw}. We work at small $g$ and fixed $g h$.  Expanding $\vec{n}$ in fluctuations using \eqref{eq_n_fluct}, we find
\begin{equation}\label{eq_S_nonlocal_2}
\begin{split}
S&\simeq\frac{1}{g^2}\int d^2y\left\{\frac{1}{2}(\pd_m \vec{\pi})^2-\frac{ g^2h^2}{2(d-2)\Omega_{d-1}}\int d^2y'\frac{\vec{\pi}(y)\cdot \vec{\pi}(y')
-[\vec{\pi}^2(y)+\vec{\pi}^2(y')]/2}{|y-y'|^{d-2}}
+O(\pi^4)\right\} \,.
\end{split}
\end{equation}
In $d<6$ the parameter $g^2 h^2$ is dimensionful and, similarly to a mass term,  it prevents the coupling $g$ from becoming strong in the IR.  Indeed from eq.~\eqref{eq_S_nonlocal_2} we see that the propagator of the Goldstone fields has a softer behaviour in momentum space for $p\rightarrow 0$ than in a local QFT
\begin{equation}\label{eq_free_ex_prop}
\langle\pi_a(y)\pi_b(0)\rangle=\int\frac{d^2p}{(2\pi)^2}\frac{e^{ipy}g^2\delta_{ab}}{p^2+\alpha_0 |p|^{d-4}}\,,
\end{equation}
where we defined the following coefficient
\begin{equation}\label{eq_alpha0}
\alpha_0=-
g^2 h^2\frac{ 2^{4-d} \pi\Gamma \left(2-\frac{d}{2}\right)}{(d-2) \Omega_{d-1} \Gamma \left(\frac{d}{2}-1\right)}
=-g^2 h^2
\frac{\Gamma \left(2-\frac{d}{2}\right)}{(4 \pi )^{\frac{d-2}{2}} }
\,,
\end{equation}
where this expression holds for $4<d<6$. Notice that $\alpha_0$ is positive.

It is now simple to check that the SSB persists at quantum level in this model. Indeed we can compute the expectation value of the order parameter similary to eq.~\eqref{eq_O_quant_pre}. Differently than in standard NLSMs, for the model at hand the self-contraction diagram is not infrared divergent
\begin{equation}
\langle\vec{\pi}^2(0)\rangle=(N-1)\int \frac{d^2p}{(2\pi)^2}\frac{1}{p^2+\alpha_0 |p|^{d-4}}=
\frac{N-1}{2\pi}\log\left(\Lambda/\alpha_0^{\frac{1}{6-d}}\right)\,.
\end{equation}
Using the wave-function renormalization in eq.~\eqref{eq_Z_O(N)},  we find the one-point function
\begin{equation}\label{eq_free_ex_n1}
\langle n_1 \rangle=1-\frac{g^2(N-1)}{2\pi}\log\left(\mu/\alpha_0^{\frac{1}{6-d}}\right)+O\left(g^4\right)\,.
\end{equation}
We see that no large logarithm appears when the renormalization scale is chosen so that $\mu\sim \alpha_{0}^{\frac{1}{6-d}}$. As promised $\alpha_0$ acts similarly to an IR regulator; however, differently than the local term \eqref{eq_L_IR}, the action \eqref{eq_S_nonlocal} preserves the full $O(N)$ symmetry. Eq.~\eqref{eq_free_ex_n1} thus represents a \emph{spontaneous} breaking.  From the long distance behavior of the propagator \eqref{eq_free_ex_prop} 
\begin{equation}
\langle\pi_a(x)\pi_b(0)\rangle\stackrel{x\rightarrow\infty}{\sim}
\frac{\delta_{ab}}{\alpha_0|x|^{6-d}}\,,
\end{equation}
it is similarly simple to check that cluster is satisfied $\lim_{x\rightarrow\infty}\langle n_A(x)n_B(0)\rangle= \delta_A^1\delta_B^1$.

For $d<6$ the RG flow never terminates in a conformal fixed-point. Indeed the one-point function of the bulk field is still given by eq.~\eqref{eq_h_1pt} to leading order in $g$.  In appendix \ref{app_current_ex} we analyze the correlation functions of the defect and bulk currents with $\vec{n}$, and show that the Ward identity~\eqref{eq_rho_int_defect} is saturated by a \emph{continuum} of states, in agreement with the general discussion in the previous section.

It is interesting to contrast these results with the behaviour of the theory in $6$ dimensions.  In this case $h$ is classically marginal and we cannot extrapolate the previous formulas. This is because $\alpha_0$ in eq.~\eqref{eq_alpha0} diverges for $d\rightarrow 6$. As we will see, this divergence signals additional contributions in the beta function~\eqref{eq_beta_O(N)} of $g$. 

It is convenient to work in a regime such that both $g$ and $g h$ are small, i.e. to take $h\sim O(1)$. This is self-consistent. Indeed it is possible to check that the coupling $h$ is marginally irrelevant, as it can be seen from the beta function
\begin{equation}\label{eq_h_defect}
\beta_h=\gamma_{\vec{n}}h=\frac{g^2(N-1)}{4\pi}h\,,
\end{equation}
where $\gamma_{\vec{n}}$ is the anomalous dimension~\eqref{eq_anomalous_O(N)}. Eq.~\eqref{eq_h_defect} can be derived requiring that the one-point function $\langle\phi_1\rangle$ in the bulk remains finite at order $O(g^2 h^2)$.

To proceed we consider the $\sim g^2 h^2$ contribution from the defect vertex to the propagator on the surface:
\begin{equation}
\begin{split}
\langle\pi_a(x)\pi_b(0)\rangle &\supset
\frac{h^2}{2}\int d^2y\int d^2z\langle\pi_a(x)\pi_b(0)
\pi_c(y)\pi_d(z)\rangle\langle\phi_c(y)\phi_d(z)\rangle
=\delta_{ab}\frac{g^4h^2}{4\pi^3}I(x)\,,
\end{split}
\end{equation} 
where
\begin{equation}\label{eq_I_div}
I(x)=\int d^2y\int d^2zG(x-y)G(z)\frac{1}{|y-z|^4}=-\frac{\pi}{2}G(x)\log(\Lambda|x|)+\text{finite}\,.
\end{equation}
Here $G(x)$ is the two-dimensional free propagator defined in eq.~\eqref{eq_prop}. In eq.~\eqref{eq_I_div} we isolated the UV divergence. To renormalize the two-point function, we modify the coupling renormalization in eq.~\eqref{eq_Z_O(N)} as $Z_g\rightarrow Z_g+\delta Z_g$, where
\begin{equation}
\delta Z_g=\frac{g^2 h^2}{16\pi^2}\log\left(\Lambda/\mu\right)\,.
\end{equation}
As a result, the beta-function of the coupling $g$ in eq.~\eqref{eq_beta_O(N)} gets modified to
\begin{equation}\label{eq_g_defect}
\beta_g=\left(\frac{h^2}{16\pi^2}-\frac{N-2}{4\pi}\right)g^3\,.
\end{equation}

From the beta functions in eq.s~\eqref{eq_h_defect} and \eqref{eq_g_defect} we can study the fate of the system at low energies.  The only fixed point of the system given by eq.s~\eqref{eq_h_defect} and \eqref{eq_g_defect} is found at $g=0$ and $h=h_*=\text{fixed}$. However no RG flow which starts from physical initial conditions $g^2(\mu_0)>0$ terminates there. This can be seen by considering the linearized solutions around such fixed-points.\footnote{The fixed-points with $h^2_*<4\pi(N-2)$, for which the right hand side of eq.~\eqref{eq_g_defect} is negative, are obviously never reached for $g^2(\mu_0)>0$. Expanding around $h^2_*>4\pi(N-2)$ and $g^2=0$, one finds that, while the NLSM coupling decreases as $g^2\sim 1/t$ for large RG time $t\sim -\log(\mu/\mu_0)$,  the fluctuation of $h$ necessarily grows logarithmically in absolute value, $\delta h=|h-h_*|\sim \log(t)$, and $h^2$ decreases until the beta function \eqref{eq_g_defect} becomes negative and $g$ starts growing. To analyze the fixed point at $h^2_*=4\pi(N-2)$ we solve for $g^2\propto-\delta h'/h_*$ from eq.~\eqref{eq_h_defect} and we find the following equation
\begin{equation}\label{eq_foot_delta_h}
\delta h''= \frac{1}{2\pi(N-1)}\delta h (\delta h')^2\,.
\end{equation}
The solution $\delta h(t)$ of eq.~\eqref{eq_foot_delta_h} can be given explicitly in terms of the inverse error function. We find that $\delta h$ never asymptotes to the fixed point $\delta h=0$ for initial boundary conditions such that $\delta h'\sim -g^2/h_*\neq 0$} We thus conclude that $h^2$ decreases in the IR and the NLSM coupling $g$ grows until it reaches the nonperturbative regime. We thus expect that the defect Goldstones acquire a mass at the strong coupling scale~\eqref{eq_Lambda_O(N)} as in the usual NLSM. 

The absence of a nontrivial interacting fixed point in six dimensions should not come as a surprise. Indeed in~\cite{Lauria:2020emq} it is shown that a free massless scalar in integer dimensions may admit nontrivial conformal defects, beyond the source defect,  only for codimension 3 and $d\geq 5$. This result implies that the model~\eqref{eq_free_ex} cannot reach an interacting conformal fixed point for $d=6$.

For completeness we also notice that close to $6$ dimensions the beta function \eqref{eq_h_defect} acquires a term $-\varepsilon/2 h$ where $\varepsilon=6-d>0$; thus the theory admits an unstable perturbative fixed point with $h^2/(4\pi)=N-2$ and $g^2/(2\pi)=(N-1)\varepsilon $. The analysis of this fixed point is analogous to that of the standard NLSM at $N=2$. In particular, the positivity of the anomalous dimension \eqref{eq_anomalous_O(N)} implies that no SSB occurs. As we increase $\varepsilon$, the coupling becomes large and we lose perturbative control over this fixed point.

Few comments are in order. First, we remark that the model~\eqref{eq_S_nonlocal} can also be regarded as a non-local theory \emph{per se}, irrespectively of the existence of bulk fields. 
A classic result from the Mermin-Wagner paper \cite{Mermin:1966fe} states that the two-dimensional long-range $O(N)$ model, which is given by eq.~\eqref{eq_S_nonlocal} without the kinetic term, cannot break spontaneously the symmetry if the nonlocal interaction term decays faster than $1/|x|^{4}$.  A similar result holds for the model \eqref{eq_S_nonlocal}. This is clear from the viewpoint of  RG: for $d>6$ the coupling $h$ is irrelevant (the interaction term decays as $1/|x|^{d-2}$) and thus at long distances the model reduces to the local NLSM. Our analysis additionally shows explicitly that SSB generically occurs for $d<6$,  since the Goldstone propagator is less singular than $1/p^2$ for $p\rightarrow 0$. Perhaps less trivially, no SSB occurs also for $d=6$, in which case $h$ is classically marginal. This is in agreement with our general discussion, since the RG flow terminates in a (trivial) DCFT in six dimensions.

The existence of surface defects spontaneously breaking a symmetry in a theory with moduli spaces is not that surprising after all. Indeed,  in the presence of flat directions in the potential it is possible to construct symmetric defects which lead to standard SSB in the bulk \cite{Prochazka:2020vog}, i.e. for which bulk operators acquire a constant expectation value.  The novelty of our setup is that the bulk field expectation value vanishes at infinite distance from the defect.  

In the future it might be possible to construct surface defects leading to SSB in different models. Our analysis suggests that suitable candidates can be found in theories with moduli spaces and \emph{fundamental} fields with sufficiently large scaling dimension (to ensure that the expectation values of bulk fields decay at large distances).\footnote{Notice that the unitarity bounds, in general, are not enough to guarantee that the expectation value of a bulk operator vanishes at large distances for a non-conformal defect; this is because conformal invariance does not fix the form of the one-point function. In the free theory example considered in this section we found that a \emph{healthy} source defect (i.e. such that $\lim_{z_{\bot}\rightarrow\infty}\langle\phi_A(z_\bot)\rangle=0$) exists only in $d>4$ (cfr. eq.~\eqref{eq_h_1pt}),  suggesting that to generalize our construction to other theories one might need to consider models whose \emph{fundamental} field has a sufficiently large scaling dimension. } Three-dimensional supersymmetric theories of the kind analyzed in \cite{Gaiotto:2018yjh,Benini:2018bhk} might be promising in this direction.

We finally remark that the model considered in this section resembles the line defect (spin impurity) in the theory of a free scalar triplet considered in \cite{Beccaria:2022bcr,Cuomo:2022xgw,Nahum:2022fqw}.  In the remaining of this section we provide some comments on symmetry breaking on line defects.

Let us consider a line defect extending in time. It is clear that in this setup there is no natural thermodynamic limit and thus all states in the Hilbert space transform in linear representations of the internal symmetry.  It is nonethelss possible to define symmetry breaking via the condition~\eqref{eq_SSB_defect}. This is equivalent to the requirement that the ground-state is degenerate and transforms in a nontrivial representation of the symmetry group. Examples of this kind are easy to construct. For instance,  one can consider a system made of $N$ qubits with trivial Hamiltonian in a empty bulk, i.e. a purely one-dimensional quantum-mechanical system; in this system the ground state transforms in the fundamental representation of the $SU(N)$ symmetry.

The model considered in \cite{Beccaria:2022bcr,Cuomo:2022xgw,Nahum:2022fqw} is similar in spirit to the trivial example of the qubit.  It consists of a free quantum-mechanics of $2s+1$ states coupled in a $SU(2)$ invaraint fashion to a free scalar triplet in the bulk. Due to the moduli space of the bulk scalar field, the interaction does not lift the degeneracy of the ground-state, similarly to the model considered in this section.
The main difference with the model analyzed in this paper is that, in the line defect considered in \cite{Beccaria:2022bcr,Cuomo:2022xgw,Nahum:2022fqw}, the ground-state transforms in a nontrivial representation independently of the number of bulk dimensions. Indeed,  for $d\geq 4$ the line defect decouples from the bulk and the Hilbert space is the tensor product of that of a spin-$s$ qubit with the Hilbert space of the free scalars. For $d<4$ instead the DQFTs considered in \cite{Beccaria:2022bcr,Cuomo:2022xgw,Nahum:2022fqw} are interacting and, similarly to the model~\eqref{eq_free_ex}, never reach a fixed-point in the infrared. (For $d$ close to $4$ there also exist conformal fixed points for which the ground state is nondegenerate.) 

Interestingly, in \cite{Cuomo:2022xgw} it was found that spin-impurities in the interacting $O(3)$ model (in $d<4$) behave very differently, and always admit a $O(3)$ symmetric ground-state for finite spin $s$. In general, we expect that line defects in CFTs without moduli spaces either admit a ground-state invariant under the internal symmetry or lead to a decoupled sector. This conclusion indeed follows by arguments analogous to those presented in sec.~\ref{sec_SSB_WI} for defects of dimension $p\geq 2$. We remark that, while for surface defects Coleman's argument rules out the existence of a decoupled defect Goldstone sector, this is not true anymore for line defects.  Technically, the reason why line defects may admit a decoupled \emph{Goldstone} sector is that, in one spacetime dimensions, it is possible to write a single-derivative kinetic term for the defect fields compatibly with all the symmetries. This is precisely the form of the action describing free qubits and leads to a propagator of the form $\sim \theta(t)+\text{const.}$, which does not violate reflection positivity nor the cluster decomposition principle.

\section{Nonlinear sigma models on conformal defects}\label{sec_log}

\subsection{Setup}\label{subsec_log_setup}

Let us suppose a $p$-dimensional defect breaks spontaneously an internal symmetry $G$ to a subgroup $H$ as in sec.~\ref{sec_SSB_WI}. We further assume that the corresponding DQFT reaches a critical fixed-point, thus excluding the possibility of a runaway flow as in the previous section.  In this case the low energy theory is naturally written as follows
\begin{equation}\label{eq_S_low_E}
S=S_{DCFT}+S_{NLSM}+S_{couple}\,.
\end{equation}
Here $S_{DCFT}$ formally represents the action for a DCFT in which the defect breaks \emph{explicitly} the symmetry group to $ H$. The details of such sector depend on the specific DQFT under consideration. Similarly to the normal boundary unvarsality class in the $O(N)$ model, this sector includes all bulk operators, which generically admit nontrivial one point functions compatibly with the unbroken group $ H$.  $S_{NLSM}$ describes a $p$-dimensional NLSM with target space $G/H$, whose generic action is written as in sec.~\ref{subsec_CMW_NLSMs}:
\begin{equation}\label{eq_NLSM_2}
S=\int d^p y\left[\frac{1}{2}\pd_m \pi^a\pd_m\pi^b g_{ab}(\pi)+\frac12 m_{IR}^2\pi^a\pi^bg_{ab}(0)+\ldots\right]\,,
\end{equation}
where we included an IR regulator and the dots stand for higher derivative terms.
The NLSM sector and the DCFT are coupled via the last term $S_{couple}$. This term restores the symmetry in the DCFT. It consists of an explicit interaction vertex between the \emph{tilt} operators of the DCFT and the Goldstone fields. To the lowest order in the fields it is written as:
\begin{equation}\label{eq_S_couple}
S_{couple}=\int d^px \hspace*{0.05em}\pi^a(x)\hat{t}_a(x)+\ldots\,.
\end{equation}
To understand the meaning of the term~\eqref{eq_S_couple}, notice that from the action~\eqref{eq_NLSM_2} we find that the defect currents for the broken generators are, to linear order,
\begin{equation}
\hat{J}^m_a=g_{ab}\pd^m\pi^b+O\left(\pi\pd\pi\right)\,.
\end{equation}
The equations of motion obtained from eq.s~\eqref{eq_NLSM_2} and~\eqref{eq_S_couple} ensure that to linear order in the fields
\begin{equation}\label{eq_low_E_EOMs}
\pd_m\hat{J}^m_a(y)=-\hat{t}_a(y)\,.
\end{equation}
Recalling that in the DCFT the tilt operator and the bulk current are related,
\begin{equation}\label{eq_bulk_tilt}
\pd_\mu J^\mu_a(x)=\delta^{d-p}(z_\bot)\hat{t}_a(y)\qquad(\text{DCFT})\,,
\end{equation}
eq.~\eqref{eq_low_E_EOMs} implies that the Ward identity for the full system~\eqref{eq_S_low_E} reads
\begin{equation}
\pd_\mu J^\mu_a(x)+\delta^{d-p}(z_\bot)\pd_m\hat{J}^m_a(y)=0\,.
\end{equation}
Therefore the coupling~\eqref{eq_S_couple} ensures that we can construct topological charges (preserved by the defect) for all the generators of the full symmetry group $G$ as in eq.~\eqref{eq_charge}. 

Notice that the simplicity of the coupling \eqref{eq_S_couple} is a consequence of the normalization we chose in eq.~\eqref{eq_bulk_tilt}. In particular, differently than in the conventions of \cite{Metlitski:2020cqy},  two-point functions of the tilt operators are not canonically normalized:
\begin{equation}\label{eq_C}
\langle \hat{t}_a(y)\hat{t}_b(0)\rangle_{DCFT}=\frac{C_{ab}}{|y|^{2p}}\,,
\end{equation}
where $C_{ab}$ is a theory dependent positive definite matrix, invariant under the unbroken group $H$. Since this matrix will be important for the analysis of two-dimensional defects, we remark that it satisfies an interesting identity~\cite{Metlitski:2020cqy,Padayasi:2021sik}.
Consider a (arbitrary) bulk operator $\mO_A(x)$ which transforms nontrivially under the broken generators. Its bulk-to-defect OPE receives contributions from both the identity and the tilt operators as
\begin{equation}
\mO_A(z_\bot,0)\sim\frac{a_A}{z_\bot^{\Delta_{\mO}}}+b_{A}^{\;a}\frac{\hat{t}_a(0)}{z_\bot^{\Delta_{\mO}-p}}+\ldots\,.
\end{equation}
One can use Ward identities to show that an infinitesimal variation $\delta_a a_A=i (Q_a^{(R)} a)_A$ generated by a broken generator $Q_a$ is related to the coefficient $b_{A}^{\;a}$ and the matrix $C_{ab}$ as
\begin{equation}
\delta_a a_A=\frac{2^{1-p} \pi ^{\frac{p+1}{2}} }{\Gamma \left(\frac{p+1}{2}\right)}b_{A}^{\;b}C_{ba}\,.
\end{equation}
The matrix $C_{ab}$ may thus be extracted from the knowledge of a sufficient number of bulk-to-defect OPE coefficients.

Higher order terms in eq.~\eqref{eq_S_couple} are constructed requiring $G$-invariance order by order in perturbation theory. We expect that the coupling~\eqref{eq_S_couple} can always be completed to ensure $G$-invariance at the nonlinear level, by supplementing it with terms of the form $\sim \pi^n\hat{t}$ with $n\geq 2$. Such nonlinearities will be irrelevant for our purposes.

\subsection{Perturbative RG flows}\label{subsec_log_RG}

For $p>2$-dimensional defects, the coupling~\eqref{eq_S_couple} is irrelevant. The canonically normalized Goldstone fields $\pi_a^{(C)}=g^{1/2}_{ab}\pi^b$ have dimension $(p-2)/2>0$, and the operator $\pi^a \hat{t}_a$ has dimension $p+(p-2)/2>p$.  Similarly, the canonically normalized defect currents $J^{(C)\,m}_{a}=\pd^m \pi_a^{(C)}$ are formally conserved in the IR, as it can be seen by sending to zero the irrelevant coupling $g^{-1}\rightarrow 0$.   Therefore the long distance description consists of a DCFT and a decoupled defect Goldstone sector.   This is in agreement with the discussion in sec.~\ref{subsec_SSB_defects} about scale-invariant fixed points and defect Goldstones.

Accordingly, the two-point functions of the currents and the defect order parameter take the form written in eq.s~\eqref{eq_J_O_g} and~\eqref{eq_J_O_g_2}. The coupling~\eqref{eq_S_couple} allows to determine the function $f_{aA}(y)$ explicitly in this case.  Since $\pd_\mu J^\mu_a(z_\bot,y)=\delta^{d-p}(z_\bot)\hat{t}_a(y)$, it is enough to compute the two-point function of the tilt operator and the defect order parameter $\hat{\mO}_A$.  The defect order parameter is represented in the IR theory in terms of the Goldstone fields as in eq.~\eqref{eq_order_parameter}.\footnote{Higher derivative operators and nontrivial DCFT operators provide additional contributions in the operator matching. Because of the unitarity bounds, these yield subleading contributions in the large distance limit of correlation functions.} Therefore, working at the first notrivial order in conformal perturbation theory and using~\eqref{eq_C}, we find
\begin{equation}\label{eq_f_calc}
\begin{split}
\frac{f_{aA}(|y|\Lambda)}{|y|^p}&=\langle \hat{t}_a(y)\hat{\mO}_A(0)\rangle 
\simeq   i\,C_{ab}g^{bc}(Q_c^{(R)} v)_A \int d^py'\frac{1}{|y-y'|^{2p}}G(y)\\
&=-\frac{C_{ab}g^{bc}\delta_c v_A}{2p(p-1)\Omega_{p-1}|y|^{2(p-1)}}\,,
\end{split}
\end{equation}
where $G(y)$ is the $p$-dimensional free propagator and we used dimensional regularization to remove a cutoff divergence.\footnote{In a cutoff scheme the result is regulated introducing a wave-function renormalization as $\hat{t}_a\rightarrow \hat{t}_a+\delta Z_{ab}\pi^b$ where $\delta Z_{ab}\propto C_{ab}\Lambda^{p}$.} Using that by dimensional analysis the Goldstone decay constant scales as $g_{ab}\sim \Lambda^{p-2}$ in terms of the cutoff scale,  we see that $f_{aA}(|y|\Lambda)\sim 1/|y\Lambda|^{p-2}$ which decreases at large distances for $p>2$ as expected.

For $p=2$-dimensional defects the interaction term is instead marginal. We can analyze the RG flow in the perturbative regime as in sec.~\ref{subsec_CMW_NLSMs}.  It is simple to check that the coupling with the tilt operator does not affect the calculation of the one-point function of the order parameter in eq.~\eqref{eq_O_quant_pre}. Thus the result for the anomalous dimensions~\eqref{eq_anomalous} remains unchanged. However, as noticed in \cite{Metlitski:2020cqy}, the coupling~\eqref{eq_S_couple} provides an additional contribution to the beta-function~\eqref{eq_beta_NLSM}. This is analogous to what we found in the model in sec.~\ref{sec_free_ex} for $d=6$, where we saw that the NLSM coupling beta function is modified to~\eqref{eq_g_defect}. However, differently than in the free theory model, a generic DCFT explicitly breaking the symmetry does not admit marginal deformations besides the tilt. This means that we do not need to consider additional running couplings besides the NLSM metric, and there is no analogue of the running of $h$ in eq.~\eqref{eq_h_defect}. In particular,  the matrix $C_{ab}$ only depends upon the way the unbroken group $H$ is embedded in $G$ and thus does not run.

To compute the change in the beta-function of the NLSM metric $g_{ab}$ with respect to local NLSMs, we notice that the Goldstones' propagator receives a one-loop contribution from the coupling~\eqref{eq_S_couple}
\begin{equation}
\begin{split}
\delta\langle\pi^a(y)\pi^b(0)\rangle& =\frac12\int d^2y\int d^2y' 
\langle\pi^a(x)\pi^b(0)\pi^c(y)\pi^d(y')\rangle_{tree}\langle \hat{t}_c(y)\hat{t}_d(z)\rangle_{DCFT} \\
&=-g^{ac}g^{bd}C_{cd}\frac{\pi}{2}\log(\Lambda |y|)G(y)+\text{finite}\,,
\end{split}
\end{equation}
where we used the result~\eqref{eq_I_div}.  For correlation functions to be finite, we need to introduce an additional contribution in the coupling renormalization. Proceeding as we did above eq.~\eqref{eq_g_defect} and recalling the result for local NLSMs~\eqref{eq_beta_NLSM}, we obtain the beta function
\begin{equation}\label{eq_modified_beta}
\beta_{ab}=\frac{\pd g_{ab}}{\pd\log\mu}=\beta_{ab}\vert_{C=0}-\frac{\pi}{2}C_{ab}=\frac{1}{2\pi}\mathcal{R}_{ab}-\frac{\pi}{2}C_{ab}+\ldots\,,
\end{equation}
where we retained only the one-loop term.

Interestingly, the beta-function~\eqref{eq_modified_beta} contains an additional \emph{negative-definite} contribution with respect to the standard NLSM result in eq.~\eqref{eq_beta_NLSM}. This term opens up the possibility of having interesting perturbative RG flows. To see this, it is convenient to decompose the metric as $g_{ab}=\hat{g}_{ab}/T$, where $\hat{g}_{ab}$ is unimodular, as we formerly did in sec.~\ref{subsec_CMW_NLSMs}. Setting $t=-\log(\mu/\mu_0)$, we find the following system of equations: 
\begin{align}\label{eq_RG_T_modified}
\frac{\pd T}{\pd t}&=\frac{T^2}{ n}\left(\frac{1}{2\pi}\hat{\mathcal{R}}-\frac{\pi}{2}\hat{C}\right)+O\left(T^3\right)\,,\\
\label{eq_RG_g_modified}
\frac{\pd \hat{g}_{ab}}{\pd t}&=T\left[\left(\frac{1}{2\pi}\hat{\mathcal{R}}_{ab}-\frac{\pi}{2}C_{ab}\right)
-\frac{1}{n}\hat{g}_{ab}\left(\frac{1}{2\pi}\hat{\mathcal{R}}-\frac{\pi}{2}\hat{C}\right)\right]+O\left(T^2\right)\,,
\end{align}
where $n=n_G-n_H$ and we defined
\begin{equation}
\hat{C}=C_{ab}\hat{g}^{ab}>0\,.
\end{equation}
For the reasons explained below eq.s~\eqref{eq_RG_T} and~\eqref{eq_RG_g}, the following condition needs to be satisfied at a fixed point\footnote{More precisely,  the beta function may or may not vanish at a fixed-point depending on the scheme (cfr. eq.~\eqref{eq_beta_v}), as we commented in footnote~\ref{foot_beta_v}. In general we should require the equality in eq.~\eqref{eq_RG_Einstein_modified} only up to a term $\nabla_a v_b+\nabla_b v_a$, where $v$ is a left-invariant vector under $G$.  Since the general argument of \cite{Polchinski:1987dy} does not apply to surface defects, this opens up the possibility of scale invariant fixed points that are not conformal, corresponding to a vector $v_a$ that is not the gradient of a scalar. We will not consider this option in the following.}
\begin{equation}\label{eq_RG_Einstein_modified}
\frac{1}{2\pi}\hat{\mathcal{R}}_{ab}-\frac{\pi}{2}C_{ab}=\frac1n\hat{g}_{ab}\left(\frac{1}{2\pi}\hat{\mathcal{R}}-\frac{\pi}{2}\hat{C}\right)\,.
\end{equation}
Eq.~\eqref{eq_RG_Einstein_modified} replaces the Einstein condition ~\eqref{eq_RG_Einstein}. It consists of $n(n+1)/2-1$ equations for the same number of variables. We thus expect that eq.~\eqref{eq_RG_Einstein_modified} generically admits one or more solutions $\hat{g}^*_{ab}$.  Notice that eq.~\eqref{eq_RG_Einstein_modified} is automatically satisfied when there is a unique rank two tensor invariant under the unbroken group $H$, in which case $\hat{g}_{ab}\propto \mathcal{R}_{ab}\propto C_{ab}$.

Given a solution of eq.~\eqref{eq_RG_Einstein_modified}, the structure of the RG flow depends on the value of the parenthesis on the right hand side~\eqref{eq_RG_T_modified}:
\begin{equation}
\alpha\equiv \frac{1}{n} \left(\frac{\pi}{2}\hat{C}-\frac{1}{2\pi}\hat{\mathcal{R}}\right)_{g=\hat{g}}\,.
\end{equation}
There are three possibilities:
\begin{itemize}
\item $\alpha<0$. In this case the coupling increases to nonperturbative values in the IR, as in local NLSMs. We thus generically expect that the Goldstones develop a gap, and the symmetry is restored.
\item $\alpha=0$. This is a non generic possibility. The fate of the RG flow in this case depends on the sign of the $O(T^3)$ term in eq.~\eqref{eq_RG_T_modified}.  When this is negative, there is an at least metastable perturbative RG flow where $T$ flows to zero.
\item $\alpha>0$. In this case there is an at least metastable RG flow to weak coupling $T\rightarrow 0$. Whether this RG flow is fully stable or not depends on the linearization of eq.~\eqref{eq_RG_g_modified} around the solution of eq.~\eqref{eq_RG_Einstein_modified}.
\end{itemize}
This situation is to be contrasted with the analysis of sec.~\ref{subsec_CMW_NLSMs}, where we found that there is no nontrivial RG flow where the coupling remains perturbative at arbitrary scales (but for the case of Abelian groups). 

In the following we analyze in detail RG flows of the third kind, for which $\alpha>0$. Assuming  that we are expanding around a stable solution $\hat{g}^{ab\,*}$ of eq.~\eqref{eq_RG_Einstein_modified} or that we fine-tuned away all dangerously relevant directions, the IR limit of the solution to eq.s~\eqref{eq_RG_T_modified} and \eqref{eq_RG_g_modified} reads
\begin{equation}\label{eq_g_RG_modified}
g^{ab}(t)\xrightarrow{t\rightarrow\infty}\frac{\tilde{g}^{ab\,*}}{t}+\ldots\,,
\quad\text{with}\;\;
\tilde{g}^{ab\,*}=\hat{g}^{ab\,*}/\alpha\,,
\end{equation}
where we retained only the leading term for $t\rightarrow\infty$.  We may now use eq.~\eqref{eq_g_RG_modified} to analyze the IR limit of correlation functions using Callan-Symanzik equation as in sec.~\ref{subsec_CMW_NLSMs}. Interestingly, eq.~\eqref{eq_g_RG_modified} exactly saturates the condition~\eqref{eq_g_condition}. Therefore, the positivity of the anomalous dimensions~\eqref{eq_anomalous} ensures that no symmetry breaking occurs even if the RG flow is perturbative.

Let us discuss this point in more detail. We plug eq.~\eqref{eq_g_RG_modified} in the formula for the anomalous dimension~\eqref{eq_anomalous} to find
\begin{equation}\label{eq_gamma_rescaled}
\gamma_{\bar{A}\bar{B}}=\frac{1}{4\pi}\frac{\tilde{g}^{ab\,*}}{t}
\left(Q_a Q_b\right)^{(R)}_{\bar{A}\bar{B}}\equiv\frac{\hat{\gamma}^*_{\bar{A}\bar{B}}}{t}\,.
\end{equation}
It is convenient to decompose the order parameter classical expectation value $\langle\hat{\mO}_{\bar{A}}\rangle=v_{\bar{A}}$ into eigenvalues of the (positive-definite) matrix $\hat{\gamma}^*_{\bar{A}\bar{B}}$:
\begin{equation}\label{eq_eigen_mod}
v_{\bar{A}}=\sum_{\alpha} \hat{c}_{\alpha} \hat{w}^{(\alpha)}_{\bar{A}}\quad
\text{such that}\quad\
\hat{\gamma}^*_{\bar{A}\bar{B}}\hat{w}^{(\alpha)}_{\bar{B}}=\hat{\bar{\gamma}}_\alpha \hat{w}^{(\alpha)}_{\bar{A}}\,,
\end{equation}
where $\hat{\bar{\gamma}}_\alpha>0$ are the eigenvalues. Solving Callan-Symanzik equation as in eq.~\eqref{eq_sol_CS}, we find the one-point function of the order parameter to logarithmic accuracy
\begin{equation}\label{eq_1pt_modified}
\langle\hat{\mO}_{\bar{A}}\rangle\sim\sum_{\alpha} \hat{c}_{\alpha} \hat{w}^{(\alpha)}_{\bar{A}}
\left(\frac{1}{\log(\mu/m_{IR})}\right)^{\hat{\bar{\gamma}}_\alpha}\xrightarrow{m_{IR}\rightarrow 0}0\,,
\end{equation}
which vanishes logarithmically as we remove the IR regulator.

The eigenvalues $\hat{\bar{\gamma}}_\alpha$ also control the long distance limit of the two-point function of the order parameter.  Consider the operators $\hat{\mO}^{(\alpha)}_A$ defined in terms of the eigenvalues \eqref{eq_eigen_mod},
\begin{equation}
\hat{\mO}^{(\alpha)}_A= \left[\Omega(\pi)\right]^{(R)}_{AB} \hat{w}^{(\alpha)}_{B}\quad\implies\quad 
\hat{\mO}_A=\left[\Omega(\pi)\right]^{(R)}_{AB} v_B=
\sum_{\alpha} \hat{c}_{\alpha}  \hat{\mO}^{(\alpha)}_A\,.
\end{equation}
By solving the Callan-Symanzik equation, we find:
\begin{equation}\label{eq_2pt_modified}
\langle(\hat{\mO}^{(\beta)})^\dagger(y)\hat{\mO}^{(\alpha)}(0)\rangle\stackrel{|y|\rightarrow\infty}{\sim}\frac{\delta_{\beta\alpha}}{\left[\log(\mu|y|)\right]^{2\hat{\bar{\gamma}}_\alpha}}\,.
\end{equation}
This result is analogous to the one in eq.~\eqref{eq_n_2pt} for the $O(N)$ model with $N<2$. 

It is useful to compare the perturbative analysis of this section with the general discussion in sec.~\ref{sec_SSB_WI}. The main point is that the coupling $g^{ab}$ is a marginally irrelevant deformation of the DCFT. Consider indeed the two-point function~\eqref{eq_f_calc} for $p=2$.  We generally expect higher order terms to introduce IR divergences. By the same comments at the end of sec.~\ref{subsec_CMW_NLSMs}, this correlator  may be nonzero in the limit $m_{IR}\rightarrow 0$ only if the product of the representation $R$ of the order parameter and that of the current, the adjoint, contains a singlet.  Even if this was the case, the results for the perturbative RG in eq.~\eqref{eq_g_RG_modified} imply that, schematically, $f_{aA}(y\Lambda)\sim \left[\log(|y|\Lambda)\right]^{-\delta}\xrightarrow{\Lambda|y|\rightarrow\infty}0$ for some $\delta>1$,\footnote{The difference $\delta-1$ is the sum of the rescaled anomalous dimensions in eq.~\eqref{eq_gamma_rescaled} and the rescaled anomalous dimension of the tilt operator $\hat{t}_a$, which is not protected in the full theory where no explicit symmetry breaking occurs. The anomalous dimension of $\hat{t}_a$, by the equations of motion, coincides with that of the defect currents (similarly to the discussion in  \cite{Skvortsov:2015pea,Giombi:2016hkj}) and it is thus positive: $\gamma_{\hat{t}}\sim g^{-1/2} C g^{-1/2}\succ 0$. Therefore the power $\delta$ is positive and larger than $1$.} as expected from a marginally irrelevant interaction.

The absence of SSB then follows from the same arguments of Coleman's theorem as explained in sec.~\ref{sec_SSB_WI}.  
\emph{Formally},  the system~\eqref{eq_S_low_E} also satisfies the requirement of describing a scale-invariant theory in the IR, if we consider all the operators of the form~\eqref{eq_order_parameter} to have dimension $\Delta=0$. Intuitively, this is because the anomalous dimension vanishes for $g^{ab}\rightarrow 0$. Accordingly, from this perspective, the logarithmic behavior of the two-point function~\eqref{eq_2pt_modified} is the typical consequence of a marginally irrelevant coupling, similarly to the two-point function of the operator $\phi^2$ in massless $\lambda\phi^4$ theory in four dimensions.

Perhaps, the most interesting observation which emerges from the analysis of the system~\eqref{eq_S_low_E} is that, despite the absence of SSB, it is possible to have weakly coupled Goldstone fields on defects for arbitrary symmetry breaking patterns.  These in turn lead to calculable logarithms inside correlation functions as in eq.~\eqref{eq_2pt_modified}. This was the main insight of~\cite{Metlitski:2020cqy}, that studied boundaries in the $O(N)$ model. Here we generalized this observation to arbitrary symmetry breaking patterns, deriving the general conditions that the low energy theory must satisfy for the Goldstone sector to remain weakly coupled along the RG (cfr. eq.~\eqref{eq_RG_Einstein_modified} and below).

\subsection{Examples}\label{subsec_log_ex}

Let us illustrate the discussion in the previous section by reviewing the example of a surface defect which \emph{classically} breaks spontaneously a $O(N)$ group down to $O(N-1)$. This was discussed first in \cite{Metlitski:2020cqy}, motivated by the application to boundaries in the $O(N)$ model \cite{PhysRevB.11.4533,PhysRevB.12.3885,Bray_1977,Ohno,McAvity:1995zd,Gliozzi:2015qsa,Giombi:2019enr,Dey:2020lwp,Giombi:2020rmc} (see also \cite{Krishnan:2023cff,Giombi:2023dqs,Raviv-Moshe:2023yvq,Trepanier:2023tvb} for the case of an interface).

The NLSM on the defect is described by the action~\eqref{eq_O(N)_NLSM} plus the coupling~\eqref{eq_S_couple}. The tilt operators $\hat{t}_a$ transform as vector under the unbroken $O(N-1)$ group and thus their two-point function is determined up to an overall factor, called $C_{\rm t}$ in~\cite{Padayasi:2021sik}:
\begin{equation}\label{eq_tt_O(N)}
\langle\hat{t}_a(y)\hat{t}_b(0)\rangle=\frac{C_{\rm t}\delta_{ab}}{|y|^4}\quad
\implies\quad
C_{ab}=C_{\rm t}\delta_{ab}\,.
\end{equation}
Using eq.~\eqref{eq_modified_beta} and the standard result \eqref{eq_beta_O(N)}, we find that the defect NLSM beta function reads
\begin{equation}\label{eq_O(N)_beta_modified}
\beta_g=\frac{\alpha}{2}g^3\,,\qquad
\alpha\equiv \frac{\pi}{2}C_{\rm t}-\frac{N-2}{2\pi}\,.
\end{equation}
The fate of the RG flow depends on the sign of $\alpha$ in eq.~\eqref{eq_O(N)_beta_modified}. Numerical results suggest that $\alpha$ is positive for $N\lesssim 5$ \cite{Padayasi:2021sik,Toldin:2021kun} for boundaries in the $3d$ $O(N)$ model (and always positive for interefaces \cite{Padayasi:2021sik}). In this case the solution to eq.~\eqref{eq_O(N)_beta_modified} is given by
\begin{equation}
g^2(\mu)=\frac{g^2(\mu_0)}{1+\alpha g^2(\mu_0)\log(\mu_0/\mu)}\,.
\end{equation}
Correlation functions behaves analogously to the standard $O(N)$ model for $1<N<2$ that we discussed in sec.~\ref{subsec_CMW_ex}. In particular the one-point and two-point functions of the order parameter read exactly as in eq.s~\eqref{eq_n_1pt} and~\eqref{eq_n_2pt} up to the replacement $q\rightarrow \frac{N-1}{2\pi\alpha}$ \cite{Metlitski:2020cqy},  in agreement with eq.~\eqref{eq_gamma_rescaled}.  Notice however that the DQFT is unitary for $N>2$. This scenario for boundaries in the $O(N)$ model is referred to as \emph{extraordinary-log} universality class.  

As another trivial example we can discuss defect NLSMs of the form $(G\times G)/G$,  whose action is given in eq.~\eqref{eq_S_GG_NLSM}. Also in this case the unbroken symmetry group implies that the tilt operator of the putative DCFT admits a diagonal two-point function as in eq.~\eqref{eq_tt_O(N)}. The beta-function~\eqref{eq_beta_GG}  is thus modified to
\begin{equation}\label{eq_ex_GG_beta}
\frac{\pd \lambda}{\pd \log\mu}=\alpha\lambda^2\,,\qquad
\alpha=\frac{\pi}{2}C_{\rm t}-\frac{C_{Adj}}{8\pi}\,.
\end{equation}
We assume $\alpha>0$ in what follows. According to the formula~\eqref{eq_anomalous}, an operator $\hat{\mO}^{(R_1\times R_2)}$ transforming in a representation $R_1\times R_2$ of $G\times G$ has anomalous dimension
\begin{equation}
\gamma_{R_1\times R_2}=\frac{\lambda}{4\pi}C_{R_1\times R_2}\equiv
\hat{\gamma}_{R_1\times R_2}\times\alpha\lambda
\,,
\end{equation}
where $C_{R_1\times R_2}=C_{R_1}C_{R_2}$ is the appropriate Casimir and we defined $\hat{\gamma}_{R_1\times R_2}$ similarly to eq.~\eqref{eq_gamma_rescaled}. We may then apply straightforwardly eq.s~\eqref{eq_1pt_modified} and \eqref{eq_2pt_modified} to compute correlation functions of the would-be defect order parameters. In particular,  the two-point function decays as (neglecting group indices)
\begin{equation}
\langle\left( \hat{\mO}^{(R_1\times R_2)} \right)^\dagger(y)\;\hat{\mO}^{(R_1\times R_2)}(0)\rangle\sim
\left[\log(\mu|y|)\right]^{-2\hat{\gamma}_{R_1\times R_2}}\,.
\end{equation}

Let us finally discuss defect NLSMs for a fully broken $SO(3)$ group.  We work in a basis of generators such that the two-point function of the tilt operator \eqref{eq_C} is diagonal:
\begin{equation}\label{eq_ex_SO3_C}
C_{ab}=\text{diag}(C_1,C_2,C_3)\,.
\end{equation}
Differently than in sec.~\ref{subsec_CMW_ex}, in this basis we cannot generally assume that the NLSM metric at the origin is diagonal if we want to preserve the simple structure of the coupling~\eqref{eq_S_couple}. This makes a general analysis rather involved. Nonetheless, as a proof of principle, we show below that there exists a fully attractive fixed-point at zero coupling when the three entries in eq.~\eqref{eq_ex_SO3_C} are almost identical. 

Notice first that, if we assume $C_{ab}=C_0\delta_{ab}$ and take the NLSM metric to be diagonal at some scale $\mu$, $g_{ab}(0)=\delta_{ab}/k_0$,  the model reduces to the $(SO(3)\times SO(3))/SO(3)$ NLSM, whose analysis we discussed above.  The parameter $\alpha$ in eq.~\eqref{eq_ex_GG_beta} in this case reads
\begin{equation}
\alpha=\frac{\pi}{2}C_0-\frac{1}{4\pi}\,.
\end{equation}

We now consider the case in which the entries of the matrix~\eqref{eq_ex_SO3_C} are almost identical to each other
\begin{equation}\label{eq_ex_C_pert}
C_1=C_0+\delta C_1\,,\quad C_2=C_0+\delta C_2\,,\quad
C_3=C_0-\delta C_1-\delta C_2\,,
\end{equation}
where  $\delta C_1\,,\delta C_2\ll C_0 \sim O(1)$.
To study this case it is natural to expand the NLSM metric around a symmetric ansatz:
\begin{equation}\label{eq_ex_g_pert}
g_{ab}(0)=\frac{1}{k_0}\left(\delta_{ab}+\delta \hat{g}_{ab}\right)\,,\qquad
\delta \hat{g}_{ab}=\begin{pmatrix}
-\delta k_1 & \hat{g}_{12} &\hat{g}_{13} \\
\hat{g}_{12} & -\delta k_2  &\hat{g}_{23} \\
\hat{g}_{13} & \hat{g}_{23} & \delta k_1+\delta k_2
\end{pmatrix}\,,
\end{equation}
where we self-consistently assume $\delta \hat{g}_{ab}\ll 1$. The parametrizations in eq. \eqref{eq_ex_C_pert} and eq.~\eqref{eq_ex_g_pert} have been chosen so that $\det(C)=C_0^3$ and $\det(g)=1/k_0^3$ to linear order in the perturbations.

We now consider the beta functions to leading order in the perturbations. We first observe that under RG flow the off-diagonal entries in eq.~\eqref{eq_ex_g_pert} tend to zero, as it follows from 
\begin{equation}
\frac{\pd \hat{g}_{ij}}{\pd\log\mu}=\hat{g}_{ij}k_0\left(\frac{1}{2\pi}+\frac{\pi}{2}C_0\right)\,\qquad (i\neq j)\,,
\end{equation}
where the parenthesis on the right hand side is always positive.
We can thus safely neglect off-diagonal components in what follows.  The beta functions of the diagonal components of the metric~\eqref{eq_ex_g_pert} to the first nontrivial order read
\begin{align}
\frac{\pd \delta k_{1/2}}{\pd\log\mu}&=k_0\left(\frac{1+\pi ^2 C_0}{2 \pi }\delta k_{1/2} +\frac{ \pi}{2}  \delta C_{1/2}\right)\,,\\
\frac{\pd k_{0}}{\pd\log\mu}&=\alpha k_0^2\,.
\end{align}
Setting $t=\log(\mu_0/\mu)$,  and given some initial conditions $k_0=k_0(\mu_0)$ and $\delta k_i=\delta k_i(\mu_0)$ at $\mu=\mu_0$, the solution of these equations is
\begin{align}
\delta k_{1/2}(\mu)&= -\frac{\pi^2\delta C_{1/2}}{1+C_0\pi^2}+\frac{1}{[1+k_0(\mu_0)\alpha t]^{\frac{1+C_0\pi^2}{2\pi\alpha}}}\left[\delta k_{1/2}(\mu_0)+\frac{\pi^2\delta C_{1/2}}{1+C_0\pi^2}\right]\xrightarrow[t \rightarrow\infty]{\alpha>0} -\frac{\pi^2\delta C_{1/2}}{1+C_0\pi^2}\,,\\
k_0(\mu)&=\frac{k_0(\mu_0)}{1+k_0(\mu_0)\alpha t}
\xrightarrow[t \rightarrow\infty]{\alpha>0} \frac{1}{\alpha t}
\,,
\end{align}
showing that the fixed point at $k_0=0$ is fully stable when $\alpha>0$.

\begin{figure}[t]
   \centering
		\subcaptionbox{  \label{fig:SO(3)_1}}
		{\includegraphics[width=0.47\textwidth]{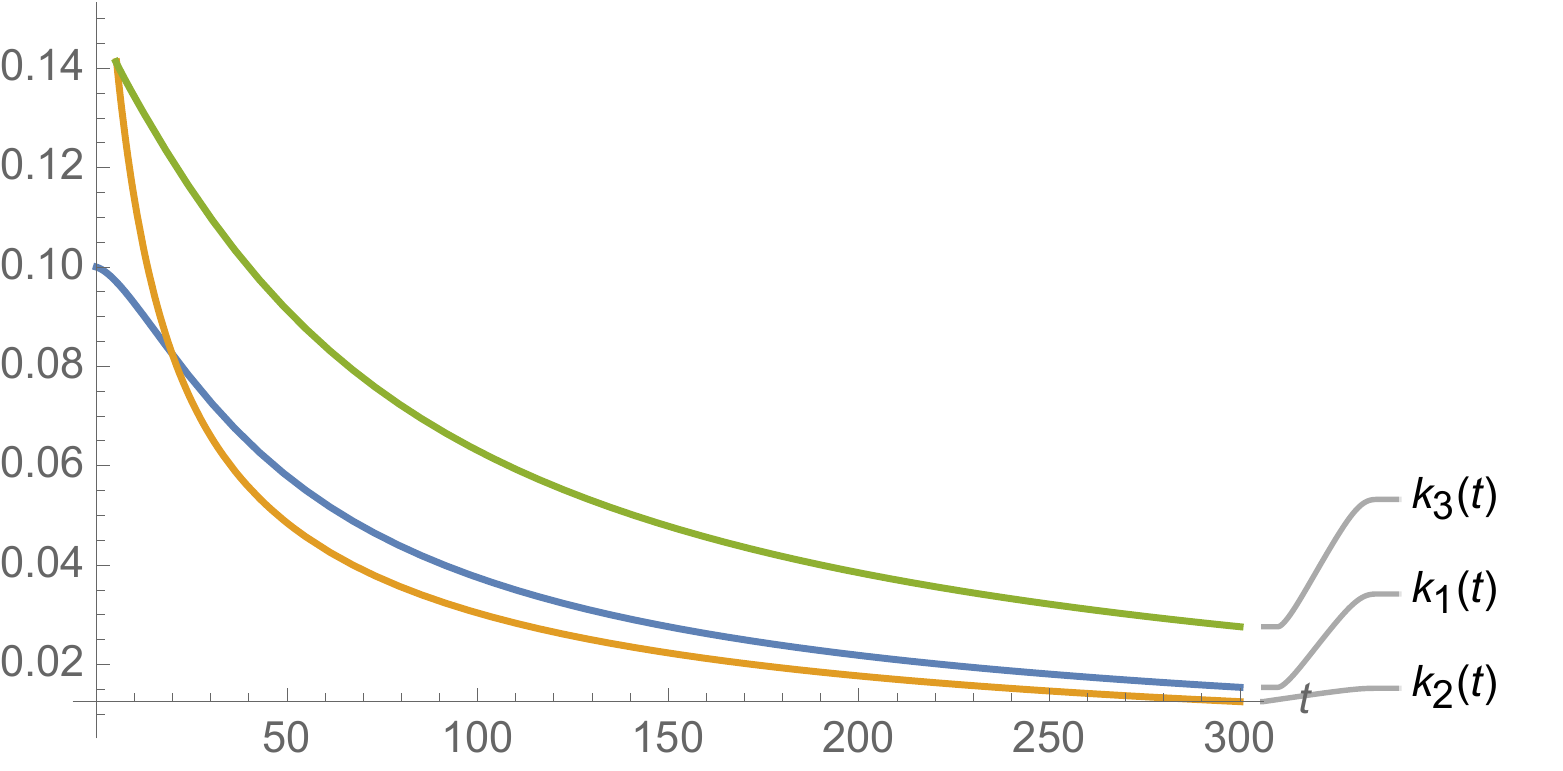}	
		}
		\subcaptionbox{ \label{fig:SO(3)_2}}
		{\includegraphics[width=0.47\textwidth]{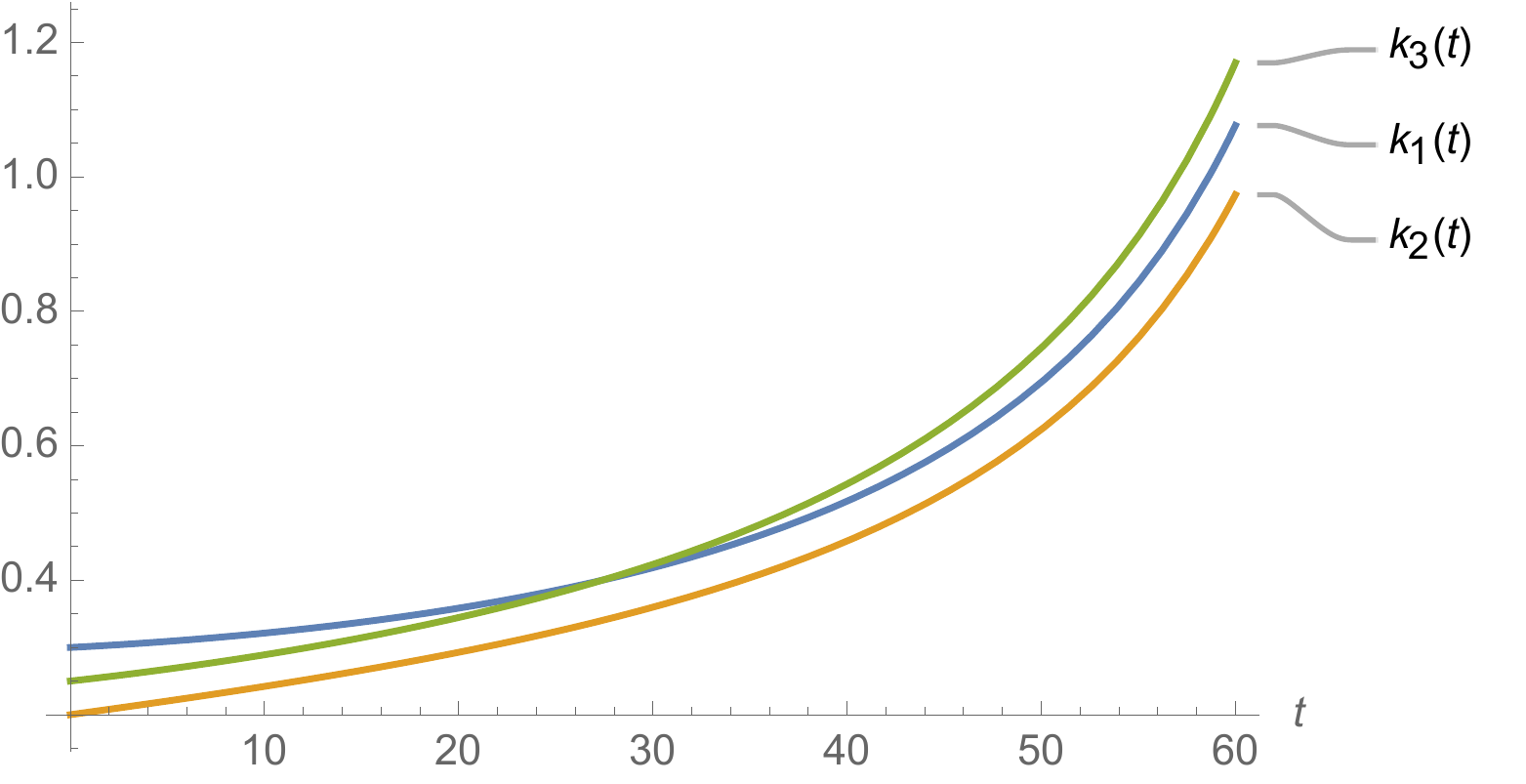}} \qquad 
\caption{Examples of the running of coupling constants in the fully broken $SO(3)$ example for a diagonal metric as in eq.~\eqref{eq_SO(3)_g}.  The horizontal axis is the RG time $t = -\log(\mu/\mu_{0})$. Fig.~\ref{fig:SO(3)_1} was obtained taking $2\pi^2 C_{ab}=\text{diag}(3.2,5.3,1.5)$ - clearly the coupling constants approach zero.  The plot in fig.~\ref{fig:SO(3)_2} corresponds to $2\pi^2 C_{ab}=\text{diag}(0.6,1.2,0.3)$ and shows an RG flow to strong coupling.}
\label{fig:SO(3)_defect}
\end{figure}
For more general values of the matrix $C_{ab}$, we observed by solving numerically the beta functions of the system that the perturbative fixed point at $g^{ab}\rightarrow 0$ is generically stable for sufficiently large values of the entries in eq.~\eqref{eq_ex_SO3_C}.  In fig.~\ref{fig:SO(3)_defect} we show two sample numerical solutions for different values of the matrix $C_{ab}$.

The main difference between the $SO(3)/\emptyset$ NLSM and the fully symmetric examples previously considered is that operators in the same representations may have different anomalous dimensions. For instance the rescaled anomalous dimension $\hat{\gamma}$ defined in eq.~\eqref{eq_gamma_rescaled} in the adjoint representation is diagonal and takes the following form
\begin{equation}\label{eq_SO(3)_anomalous}
\hat{\gamma}_{fund.}=\frac{1}{2\pi\alpha}\text{diag}\left(1
+\frac{\pi^2\delta C_1}{2(1+C_0\pi^2)}
,1+\frac{\pi^2\delta C_2}{2(1+C_0\pi^2)},1-
\frac{\pi^2(\delta C_1+\delta C_2)}{2(1+C_0\pi^2)}
\right)\,.
\end{equation}
To be concrete, consider an operator defined through the coset matrix in the adjoint as
\begin{equation}
\hat{\mO}_a(y)=\left[\Omega(\pi)\right]_{a 1}\,.
\end{equation}
By the discussion around eq.~\eqref{eq_2pt_modified}, the anomalous dimension matrix~\eqref{eq_SO(3)_anomalous} implies that at long distances its two-point function reads
\begin{equation}
\langle\hat{\mO}_b(y) \hat{\mO}_a(0)\rangle\sim \delta_{ab}
\left[\log(\mu|y|)\right]^{-2(\hat{\gamma}_{fund.})_{11}}\,,
\end{equation}
where $(\hat{\gamma}_{fund.})_{11}$ is the first entry on the right-hand side of eq.~\eqref{eq_SO(3)_anomalous}.

\section*{Acknowledgements}

We thank P. Ferrero, A.Gimenez-Grau, P.Kravchuk, A. Podo and S.Zhong for useful discussions.  We are particularly grateful to M.~Mezei for collaboration at the early stages of this project and to Z.~Komargodski, M.~Metlitski,  and A. Raviv-Moshe for valuable comments on a preliminary version of this manuscript.  We also thank M. Meineri for carefully reviewing this manuscript, providing several useful comments.
GC is supported by the Simons Foundation (Simons Collaboration on the Non-perturbative Bootstrap) grants 488647 and 397411.

\appendix

\section{Details on the free scalar example}

\subsection{Partition function and runaway RG flow}\label{app_runaway}

We consider the DQFT obtained setting $\hat{n}_A=\delta_A^1$ in eq.~\eqref{eq_free_ex}. We can obviously neglect the field components with $A>1$ in eq.~\eqref{eq_free_ex}, and the action reduces to
\begin{equation}\label{eq_ex_free_app}
S=\frac{1}{2}\int d^dx(\pd\phi)^2-h\int_{\Sigma} d^2\sigma\sqrt{G(\sigma)}\,\phi(x(\sigma))\,,
\end{equation}
where $\Sigma$ is an arbitrary two-dimensional surface with metric $G_{\alpha\beta}$.  The coupling $h$ is relevant for $d<6$, marginal in $d=6$, irrelevant otherwise. We would like to compute the universal part of the partition function for a spherical defect placed on a sphere $\Sigma=S^2$.  This quantity decreases between fixed-points of the RG flow in \cite{Jensen:2015swa} (see also \cite{Shachar:2022fqk}). At the fixed-points, the partition function is related to the coefficient $b$ of the conformal anomaly \cite{Jensen:2015swa}. 

In $d>2$ the equations of motion is solved by
\begin{equation}
\phi_{cl}(x)=\frac{h}{(d-2)\Omega_{d-1}}\int_{\Sigma} d^2\sigma\sqrt{G(\sigma)}\frac{1}{|x-x(\sigma)|^{d-2}}\,.
\end{equation}
The fluctuations around $\phi_{cl}$ are insensitive to the existence of the defect. Therefore, the defect partition function for a spherical surface defect of radius $R$, normalized by the partition function of the theory without the defect,  is obtained by simply plugging the classical solution back into the action. We find
\begin{equation}\label{eq_Z_free}
\begin{split}
\log(Z_{defect}/Z_{bulk}) &=\frac{2\pi h^2 R^{6-d}}{(d-2)\Omega_{d-1}}\int_{0}^\pi d\theta\sin\theta\int_0^{2\pi} d\phi\frac{1}{\left[4\sin^2(\theta/2)\right]^{\frac{d-2}{2}}} \\
&=
-R^{6-d}2^{3-d} \pi ^{2-\frac{d}{2}} \Gamma \left(\frac{d}{2}-2\right)\,,
\end{split}
\end{equation}
where the integral was evaluated in dimensional regularization. The pole for $d\rightarrow 4$ can be renormalized by a cosmological constant counterterm on the defect $\int d^2\sigma\sqrt{G}\sim R^2$, since this divergence is quadratic in $R$. Additionally, we can always add a Ricci scalar counterterm $\int d^2\sigma\sqrt{G}\mathcal{R}\sim R^0$. The result \eqref{eq_Z_free} in $d=6$ is thus a pure counterterm as expected by the marginality of the coupling. 

\begin{figure}[t!]
\centering
\includegraphics[scale=0.7]{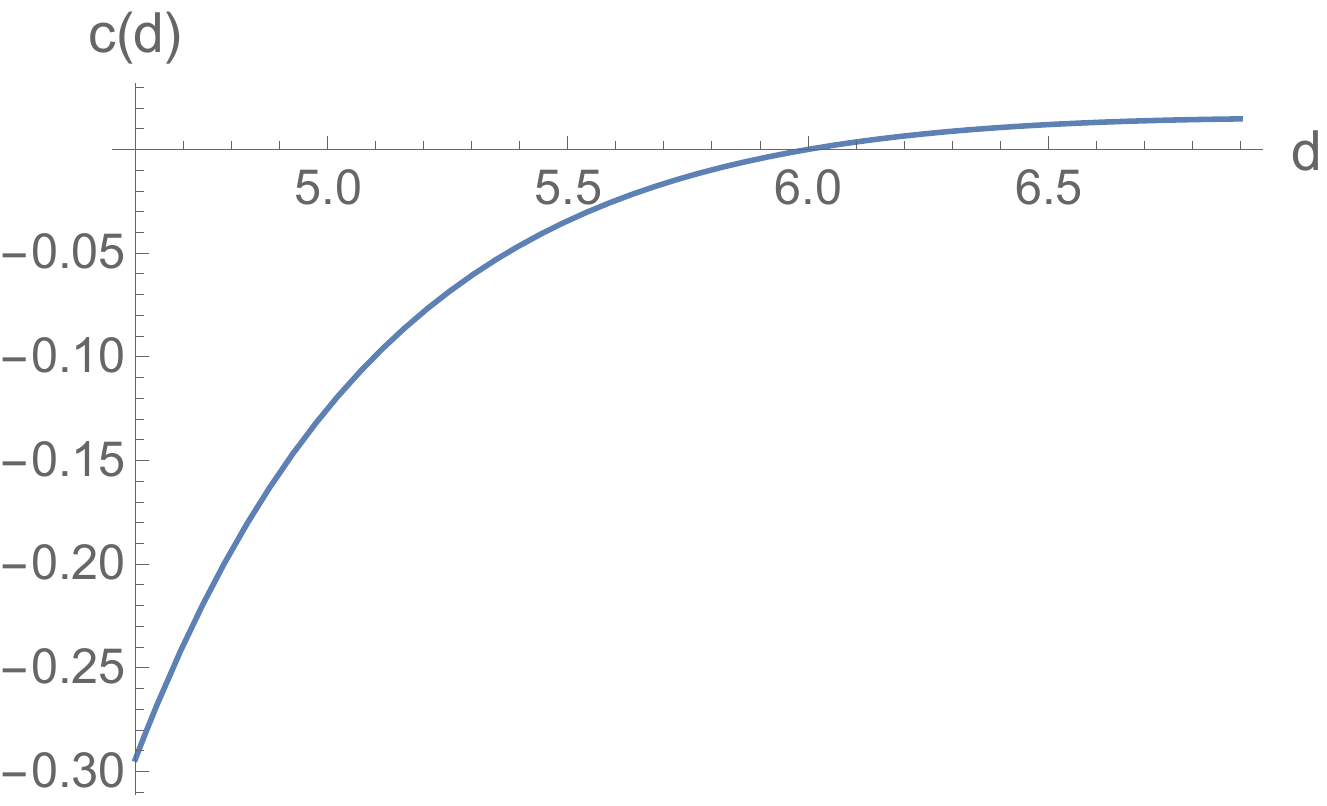}
\caption{Plot of the function $c(d)$ defined in eq.~\eqref{eq_cd}}
\label{fig:cd}
\end{figure}
To obtain a scheme-independent quantity we consider the renormalized defect entropy as defined in \cite{Shachar:2022fqk}
\begin{equation}\label{eq_free_b}
s(R)=\frac{1}{2}\left(R\pd_R-R^2\pd_R^2\right)\log(Z_{defect}/Z_{bulk})
=h^2 R^{6-d} c(d)\,,
\end{equation}
where we defined
\begin{equation}\label{eq_cd}
c(d)=2^{3-d} \pi^{2-\frac{d}{2}} \Gamma \left(\frac{d}{2}-1\right)\,.
\end{equation}
The function $c(d)$ is shown in fig.~\ref{fig:cd}. It is negative for $d<6$, while it vanishes in $d=6$, and it is positive for $d>6$. For instance in $d=5$ we find $s=-R/8$ and we see that the renormalized defect entropy vanishes in the ultraviolet ($R\rightarrow 0$) while $s\rightarrow -\infty$ in the infrared for $R\rightarrow \infty$.  The RG flow thus never terminates in an infrared DCFT for $d<6$. As explained in sec.~\ref{sec_free_ex}, we expect this behaviour to be possible only in theories with a moduli space of vacua. 

It is interesting to check the consistency of the result \eqref{eq_free_b} with the sum rule derived in \cite{Shachar:2022fqk}, which reads
\begin{equation}\label{eq_s_thm}
s(R)=s_{UV}-\frac{1}{2}\int_{S^2} d^2\sigma_1\sqrt{G(\sigma_1)}
\int_{S^2} d^2\sigma_2\sqrt{G(\sigma_2)}\langle\hat{T}(\sigma_1)\hat{T}(\sigma_2)\rangle_c\left[1-\hat{n}(\sigma_1)\cdot\hat{n}(\sigma_2)\right]\,,
\end{equation}
where $\langle\hat{T}(\sigma_1)\hat{T}(\sigma_2)\rangle_c$ is connected two-point function of the defect stress-tensor, and $\hat{n}(\sigma)$ is a unit vector on $S^2$, such that $2[1-\hat{n}(\sigma_1)\cdot\hat{n}(\sigma_2)]$ yields the chordal distance on the sphere between the points $\sigma_1$ and $\sigma_2$.  $s_{UV}$ is the renormalized defect entropy at the UV fixed-point. In the model~\eqref{eq_ex_free_app} we thus have $s_{UV}=0$, while the trace of the defect stress tensor is given by
\begin{equation}\label{eq_T_thm}
\hat{T}=\beta_h\phi(x(\sigma))\quad
\implies
\langle\hat{T}(\sigma_1)\hat{T}(\sigma_2)\rangle_c=
\frac{\beta_h^2}{(d-2)\Omega_{d-1}\left[2-2\hat{n}(\sigma_1)\cdot\hat{n}(\sigma_2)\right]^{\frac{d-2}{2}}}\,,
\end{equation}
where $\beta_h=\frac{d-6}{2}h$ is the beta function of $h$. From eq.~\eqref{eq_T_thm} we find
\begin{equation}
\begin{split}
\int_{S^2} d^2\sigma_1\sqrt{G(\sigma_1)}&
\int_{S^2} d^2\sigma_2\sqrt{G(\sigma_2)}\langle\hat{T}(\sigma_1)\hat{T}(\sigma_2)\rangle_c\left[1-\hat{n}(\sigma_1)\cdot\hat{n}(\sigma_2)\right]\\
&=
\frac{8\pi^2 R^{6-d}}{(d-2)\Omega_{d-1}}\int_0^\pi d\theta\frac{\sin\theta\left(1-\cos\theta\right)}{\left[4 \sin ^2\left(\frac{\theta }{2}\right)\right]^{\frac{d-2}{2}}}=
\frac{R^{6-d} \Gamma \left(\frac{d}{2}-1\right)}{(6-d)2^{d-6}\pi ^{\frac{d}{2}-2} }\,,
\end{split}
\end{equation}
from which it is straightforward to verify eq.~\eqref{eq_s_thm}.

The same calculation yields the leading order result for $s$ in the model~\eqref{eq_free_ex}, where the defect Goldstone's fields are dynamical, for $d<6$. In that model there are also additional subleading contributions proportional to the coupling $g^2$ evaluated at the scale $\mu\sim \alpha_0^{\frac{1}{6-d}}$, where $\alpha_0$ is given in eq.~\eqref{eq_alpha0}.

Finally we remark that one can obviously construct analogous examples of free theory source defects of different dimensions. The analysis in the case of a line defect was presented in \cite{Cuomo:2021kfm}, and examples where such source line defects are coupled to additional degrees of freedom were considered in \cite{Cuomo:2022xgw}. In these examples the trace of the defect stress tensor does not vanish in the infrared, and the defect entropy\footnote{See \cite{Cuomo:2021rkm} for the definition in the case of line defects.} $s$ approaches $-\infty$ for $R\rightarrow-\infty$. These two facts are not independent, since the sum rules of \cite{Cuomo:2021rkm} (for line defects) and \cite{Shachar:2022fqk} (eq.~\eqref{eq_s_thm}) relate the two-point of the trace of the defect stress tensor with $s$. To the best of our knowledge, the runaway RG flows in free theory considered here and in \cite{Cuomo:2021kfm,Cuomo:2022xgw} are the only known examples of defect RG flows which do not lead to DCFTs in the infrared (for unitary DQFTs).

\subsection{The Ward identity}\label{app_current_ex}

The bulk and defect currents of the model \eqref{eq_free_ex} are given by, respectively:
\begin{align}
J^\mu_{AB} &=\phi_A\pd^\mu\phi_B-\phi_B\pd^\mu\phi_A\,, \\
\hat{J}^m_{AB} &=\frac{1}{g^2}\left(n_A\pd^m n_B-n_B\pd^m n_A\right)\,.
\end{align}
We show below that the correlators $\langle J_{AB}^\mu(x) n_C(0)\rangle$ and $\langle \hat{J}_{AB}^m(y) n_C(0)\rangle$ satisfy the Ward identity \eqref{eq_defect_pole} without the existence of defect Goldstone bosons. 
Let us assume that the order parameter is aligned in the first direction $\langle n_A\rangle=\delta_A^1$.  We denote with $a=2,\ldots, N$ the unbroken indices. The nontrivial contribution to the Ward identity arises from the components $n_a=\pi_a$ of the defect field and $J^m_{1a}$ and $\hat{J}^m_{1a}$ of the currents. Explicitly, the Ward identity reads
\begin{align}\label{eq_app_Ward}
    \langle \left(\partial_{m}J^{m}_{1a}(y,z_{\perp}) + \delta^{d-2}(z_{\perp})\partial_{m}\hat{J}^{m}_{1b}(y)\right)n_{b}(0)\rangle = -\delta_{ab}\langle n_{1} \rangle \delta^2(y)\delta^{d-2}(z_{\perp})\,,
\end{align}
A simple calculations shows that to leading order
\begin{align}\label{eq_J_n_b}
\int d^2y \,e^{iky}\langle J^m_{1a}(y,z_\bot) n_b(0)\rangle &= i
\frac{2^{\frac{d}{2}-3} \Gamma \left(\frac{d}{2}-2\right)}{\pi ^{\frac{d}{2}-1} \Gamma \left(2-\frac{d}{2}\right)}
\frac{\alpha_0|z_{\perp}|^{6-\frac{3d}{2}}k^{m}}{|k|^{4-\frac{d}{2}}+\alpha_{0}|k|^{\frac{d}{2}-2}}K_{\frac{d}{2}-2}(|k||z_{\perp}|)\delta_{ab}  \,,\\
\label{eq_J_n_d}
\int d^2y \,e^{iky}\langle \hat{J}^m_{1a}(y) n_b(0)\rangle &=   -i\frac{k^{m}}{|k|^{2}+\alpha_{0}|k|^{d-4}} \delta_{ab}\,.
\end{align}
From these expressions we can extract the spectral densities. Recalling the definitions
\begin{align}\label{eq_app_spectral_1}
    &\langle J^m_{1a}(k,z_\bot) n_b(-k)\rangle \equiv ik^m \delta_{ab} F_{bulk}(k^{2},z_\perp)= k^m \delta_{ab}\int_{0}^{\infty} d\mu^{2}\frac{\rho_{bulk}(\mu^{2},z_{\perp})}{k^{2}+\mu^{2}}\,,\\
    &\langle \hat{J}^m_{1a}(k) n_b(-k)\rangle \equiv ik^m \delta_{ab} \hat{F}_{P}(k^{2})= k^m \delta_{ab}\int_{0}^{\infty} d\mu^{2}\frac{\hat{\rho}_{P}(\mu^{2})}{k^{2}+\mu^{2}}\,,\label{eq_app_spectral_2}
\end{align}
the Ward identity~\eqref{eq_app_Ward} implies
\begin{align}\label{eq_app_rho_WI}
    \int d^{d-2}z_{\perp} \rho_{bulk}(\mu^{2},z_{\perp}) + \hat{\rho}_{P}(\mu^{2}) = -i\langle n_{1}\rangle \delta(\mu^{2})\,.
\end{align}

The spectral density is extracted upon taking the discontinuity of the correlators in eq.s~\eqref{eq_J_n_b} and~\eqref{eq_J_n_d}:
\begin{align}
    \rho(\mu^{2}) = -\frac{i}{\pi} \lim_{\epsilon\rightarrow 0^+}\text{Im}[F(\mu^{2}e^{i(\pi-\epsilon)})]\,.
\end{align}
Following this procedure, we obtain the following bulk and defect spectral densities
\begin{align}\label{eq_free_ex_rho_bulk}
\rho_{bulk}(\mu^2,z_\bot) =  
&-i \frac{2^{\frac{d}{2}-3} \Gamma \left(\frac{d}{2}-2\right)}{\pi ^{\frac{d}{2}-1} \Gamma \left(2-\frac{d}{2}\right)}
\frac{\alpha_0|z_{\perp}|^{6-\frac{3d}{2}}} {\mu^{2 - \frac{d}{2}}}
\left\{\frac{\alpha_{0}\mu^{d-4} \sin(\frac{\pi}{2}d) Y_{\frac{d}{2}-2}(\mu |z_{\perp}|)}{\mu^{4} + \alpha_{0}^{2}\mu^{2(d-4)}- 2\alpha_{0}\mu^{d-2}\cos(\frac{\pi}{2}d)} \right. \nonumber \\
&\left.\hspace*{12em}+\frac{\left[\mu^{2} - \alpha_{0}\mu^{d-4}\cos(\frac{\pi}{2}d)\right] J_{\frac{d}{2}-2}(\mu |z_{\perp}|)  }{\mu^{4} + \alpha_{0}^{2}\mu^{2(d-4)}- 2\alpha_{0}\mu^{d-2}\cos(\frac{\pi}{2}d)}\right\}\,,\\
\hat{\rho}_P(\mu^2) =& - \frac{i}{\pi}\frac{ \alpha_{0} \sin(\frac{\pi}{2}d)\mu^{d-4}}{\mu^{4} + \alpha_{0}^{2}\mu^{2(d-4)} - 2\alpha_{0}\mu^{d-2}\cos(\frac{\pi}{2}d)}\,. 
\label{eq_free_ex_rho_defect}
\end{align}
To obtain these formulas we used the following identity to extract the imaginary part of the Bessel function
\begin{align}
     K_{\alpha}(x) = e^{-i\frac{\pi}{2}(\alpha+1)}\left[J_{\alpha}(-ix) - i Y_{\alpha}(-ix)\right]\,.
\end{align}
Eq.s~\eqref{eq_free_ex_rho_bulk} and \eqref{eq_free_ex_rho_defect} show that the Ward identity is completely saturated by the contribution of a continuum of states to the spectral density.  

Let us now verify the Ward identity~\eqref{eq_rho+rho_defect} explicitly. To this aim we simply need to integrate eq.~\eqref{eq_free_ex_rho_bulk}. The integration can be readily performed in the spherical coordinates. For the first term in $\rho_{bulk}$, involving $Y_{\frac{d}{2}-2}(\mu|z_{\perp}|)$, the relevant integral reads
\begin{align}
    \int d^{d-2}z_{\perp}|z_{\perp}|^{6-\frac{3d}{2}}Y_{\frac{d}{2}-2}(\mu |z_{\perp}|) 
     &= \Omega_{d-3}\int_{0}^{\infty} dz z^{-\frac{d}{2} + 3}Y_{\frac{d}{2}-2}(\mu z) \nonumber \\
     & =\mu^{\frac{d}{2}-4}\frac{2^{4-\frac{d}{2}}\pi^{\frac{d}{2}-1}}{\Gamma(\frac{d}{2}-1)\Gamma(\frac{d}{2}-2)}\cot(\frac{\pi}{2}d)\,.
\end{align}
To integrate the second term in $\rho_{bulk}$, which involves $J_{\frac{d}{2}-2}(\mu|z_{\perp}|)$, it is convenient to use the integral representation of the Bessel function:
\begin{align}
     \int d^{d-2}z_{\perp}|z_{\perp}|^{6-\frac{3d}{2}}J_{\frac{d}{2}-2}(\mu |z_{\perp}|) 
    &= \Omega_{d-3}\int_{0}^{\infty} dz z^{-\frac{d}{2} + 3}J_{\frac{d}{2}-2}(\mu z) \nonumber \\
    & = \frac{2^{3-\frac{d}{2}}\Omega_{d-3}\mu^{\frac{d}{2}-2}}{\sqrt{\pi}\Gamma(\frac{d+1}{2}-2)}\int_{0}^{1}dt\int_{0}^{\infty} dz \, z (1-t^{2})^{\frac{d-1}{2}-2}\cos(\mu zt) \nonumber \\
    & = \frac{2^{3-\frac{d}{2}}(d-5)\Omega_{d-3}\mu^{\frac{d}{2}-3}}{\sqrt{\pi}\Gamma(\frac{d+1}{2}-2)}
    \int_{0}^{1}dt \,(1-t^{2})^{\frac{d-1}{2}-3} \int_{0}^{\infty} dz \, \sin(\mu z) \nonumber \\
    & = \frac{2^{4-\frac{d}{2}}\pi^{\frac{d}{2}-1}\mu^{\frac{d}{2}-3} }{\Gamma(\frac{d}{2}-2)\Gamma(\frac{d}{2}-1)}\left[\pi \delta(\mu) + \frac{1}{\mu}\right]\,.
\end{align}
In the third line we integrated by parts on $t$, and re-scaled the integration variable $z $ by $t$. Using these results, we readily obtain the integrated spectral density
\begin{equation}\label{eq_app_rho_int}
\int d^{d-2}z_\bot \rho_{bulk}(\mu^2,z_\bot) = \frac{i}{\pi}\frac{ \alpha_{0} \sin(\frac{\pi}{2}d)\mu^{d-4}}{\mu^{4} + \alpha_{0}^{2}\mu^{2(d-4)} - 2\alpha_{0}\mu^{d-2}\cos(\frac{\pi}{2}d)} -i \delta(\mu^{2})\,.
\end{equation}
Here we used the following identity to rewrite the delta function
\begin{align}\label{eq_app_delta}
\delta(\mu)=\lim_{m^2\rightarrow 0^+}\mu\,\delta(\mu^2-m^2)\left[\theta(\mu)-\theta(-\mu)\right]\equiv \mu\, \delta(\mu^2)\,,
\end{align}
where it is important that the limit $m^2\rightarrow 0^+$ is take from above since the change of variables from $\mu$ to $\mu^2$ is singular at $\mu=0$. The limit in eq.~\eqref{eq_app_delta} provides the suitable definition for $\delta(\mu^2)$ in the spectral decompositions~\eqref{eq_app_spectral_1} and \eqref{eq_app_spectral_2}.

In conclusion, we showed that the delta-function contribution in the Ward identity arises upon integration over $z_\bot$ of the bulk spectral density, in agreement with the discussion below eq.~\eqref{eq_rho_int_defect}. It can be checked that the result~\eqref{eq_app_rho_int} summed to the defect spectral density~\eqref{eq_free_ex_rho_defect}, reproduces the expected Ward identity~\eqref{eq_app_rho_WI} using that $\langle n_1\rangle= 1$ to the order of interest.

\section{Source defects and moduli spaces}\label{app_source_defects}
	
In this appendix we clarify the relation between moduli spaces and source defects of the sort considered in sec.~\ref{sec_free_ex}.

As noted in \cite{Cuomo:2021kfm}, some theories admit $p$-dimensional defects that never become conformal at large distances.  This means that, at large distances, one-point functions take the form 
\begin{equation}\label{eq_app_source_nc}
\langle\mO_{\Delta,X}(z_\bot)\rangle\sim\frac{\mu^\delta}{|z_{\bot}|^{\Delta-\delta}}\,,
\end{equation}
where $|z_\bot|$ is the distance from the defect, $\delta>0$ and $\mu$ is some mass scale, corresponding to a defect relevant coupling.  Since conformal invariance of the bulk theory allows choosing the dimensionful defect coupling $\mu$ at will,  we can make the expectation value of the field arbitrarily large at long distances.  Therefore we expect that the CFT admits a flat direction, i.e. a moduli spaces of vacua where the conformal symmetry is spontaneously broken. The simplest example of this kind is the defect obtained adding to the action of a free scalar in the $d$-dimensions the term $h\int d^py \,\phi$ with $p>(d-2)/2$.

Conversely, given a CFT with a moduli space, we can easily prove that defects of the sort~\eqref{eq_app_source_nc} exist - at least if we allow for fractional codimensions. The argument uses the effective theory on the  moduli space. Such effective theory always includes a dilaton $\phi$ as well as other fields which couples to $\phi$ through irrelevant terms (in the sense of RG): $S=\frac12\int d^dx(\pd\phi)^2+\ldots$.  In general the effective theory is holds around the vacuum $\langle\phi\rangle=v$, but it can also be trusted around other backgrounds as long as higher derivative corrections are suppressed, i.e. $\pd^2/|\langle\phi\rangle|^{\frac{4}{d-2}}\ll 1 $ schematically.  In particular, a solution of the equations of motion is given by 
\begin{equation}
\phi=\frac{\mu^{\frac{d}{2}-p-1}}{|z_\bot|^{d-p-2}}\,.
\end{equation}
Higher derivative corrections are suppressed by relative factors of $\sim (z_{\bot}\mu)^{-4\left(\frac{p}{d-2}-1/2\right)}$ and are therefore small for $p>\frac{d-2}{2}$ and $z\gg 1/\mu$, making the solution trustworthy at large distances. This solution corresponds to a non-conformal $p$-dimensional source defect. 

Let us now discuss the case of a conformal source defect. Namely, we suppose now that the CFT under consideration admits a $p$-dimensional \emph{conformal} defect with a non-compact marginal parameter $h \in \mathds{R}$ such that one-point functions take the form
\begin{equation}\label{eq_1pt_foot}
\langle\mO_{\Delta,X}(z_\bot)\rangle=\xi_{\mO}\frac{h}{|z_{\bot}|^{\Delta}}\,,
\end{equation}
where $\xi_{\mO}$ are theory dependent numbers. An example of this kind is the defect $h\int d^py \,\phi$ for $p=(d-2)/2$ in a free scalar theory, as we saw in sec.~\ref{sec_free_ex} for $p=2$ and $d=6$. In general,  we expect that the existence of a conformal defect of the sort~\eqref{eq_1pt_foot} also implies the existence of a moduli space. To see this, let us set $h=|\bar{z}_{\bot}|^{\Delta}v^{\Delta}$ and  $z_\bot=\bar{z}_\bot+\delta z_{\bot}$, such that for $\bar{z}_\bot\gg \delta z_{\bot}$ eq.~\eqref{eq_1pt_foot} takes the form
\begin{equation}\label{eq_1pt_foot2}
\langle\mO_{\Delta,X}(z_\bot)\rangle=\xi_{\mO}v^{\Delta}\left[1+O\left(
\frac{\delta z_\bot}{\bar{z}_\bot}\right)\right]\,.
\end{equation}
Since by assumption $h \in \mathds{R}$ is a marginal parameter, we can tune $v$ arbitrarily, therefore modifying the VEV of the operator at arbitrary large distances from the defect. Obviously, this is possible only if the CFT admits a moduli space of vacua. Notice that eq.~\eqref{eq_1pt_foot} corresponds to a constant VEV for the theory rescaled to AdS$_{p+1}\times S^{d-p-1}$  from this viewpoint the limit in~\eqref{eq_1pt_foot2} simply states that, zooming at distances much smaller than the AdS radius around a vacuum with constant VEV for the fields, we recover the flat space physics of the moduli space. 

It is important to remark that, differently from the non-conformal defect discussed before, we do not expect conformal defects of the sort~\eqref{eq_1pt_foot} to exist in a generic CFT with a moduli space. Indeed, from the viewpoint of the moduli space EFT, \eqref{eq_1pt_foot} corresponds to a solution for the dilaton $\phi=h/|z_\bot|^{\frac{d-2}{2}}$. This is a solution of the leading order EOMs only for $p=\frac{d-2}{2}$. However, higher derivative EFT terms in general do not vanish nor are suppressed on this solution, and thus create a potential for $h$ \cite{Hinterbichler:2022ids} - whose minimum in general does not lie within EFT.  It would be interesting to study whether such conformal source defects exists in some interacting supersymmetric theories with moduli spaces.

\bibliography{Biblio}
	\bibliographystyle{JHEP.bst}

\end{document}